\documentclass[preprint]{aastex631}

\newcommand{\noisechisel}{\textit{NoiseChisel}}
\newcommand{\gnuastro}{\textit{Gnuastro}}

\newcommand{\magd}{\rm{mag\,d^{-1}}}

\newcommand{\trimfig}[4]{\hfill\vbox{\parskip=0pt\hsize=#2
\includegraphics[width=#2, trim=#3, clip]{#1}\vskip2pt\vtop{\centering
\footnotesize
\hsize=#2
#4\vskip1pt
}}\hfill}

\shorttitle{The optical monitoring of the Didymos-Dimorphos with DK}
\shortauthors{Ro{\.z}ek et al.}
\graphicspath{{./}{figures/}}

\begin{document}

\title{Optical monitoring of the Didymos-Dimorphos asteroid system with the Danish telescope around the DART mission impact}

\correspondingauthor{Agata Ro{\.z}ek}
\email{a.rozek@ed.ac.uk}

\author[0000-0003-2341-2238]{Agata Ro{\.z}ek}
\affiliation{Institute for Astronomy, University of Edinburgh, Royal Observatory, Edinburgh, EH9 3HJ, UK}

\author[0000-0001-9328-2905]{Colin Snodgrass}
\affiliation{Institute for Astronomy, University of Edinburgh, Royal Observatory, Edinburgh, EH9 3HJ, UK}

\author[0000-0001-7303-914X]{Uffe G. J{\o}rgensen}
\affiliation{Centre for ExoLife Sciences, Niels Bohr Institute, University of Copenhagen, {\O}ster Voldgade 5, 1350 Copenhagen,  Denmark}

\author[0000-0001-8434-9776]{Petr Pravec}
\affiliation{Astronomical Institute of the Academy of Sciences of the Czech Republic, Fri\v{c}ova 298, Ond\v{r}ejov, CZ-25165, Czech Republic}

\author[0000-0002-7520-8389]{Mariangela Bonavita}
\affiliation{Institute for Astronomy, University of Edinburgh, Royal Observatory, Edinburgh, EH9 3HJ, UK}

\author[0000-0003-2935-7196]{Markus Rabus} 
\affiliation{Departamento de Matem\'atica y F\'isica Aplicadas, Facultad de Ingenier\'ia,  Universidad Cat\'olica de la Sant\'isima Concepci\'on, Alonso de Rivera 2850, Concepci\'on, Chile}

\author[0000-0001-5098-4165]{Elahe Khalouei} 
\affiliation{Astronomy Research Center, Research Institute of Basic Sciences, Seoul National University, 1 Gwanak-ro, Gwanak-gu, Seoul 08826, Korea}

\author[0000-0001-9330-5003]{Pen{\'e}lope Longa-Pe{\~n}a}
\affiliation{Centro de Astronom{\'{\i}}a, Universidad de Antofagasta, Av.\ Angamos 601, Antofagasta, Chile}

\author[0000-0002-5854-4217]{Martin J. Burgdorf}
\affiliation{Universit{\"a}t Hamburg,  Faculty of Mathematics, Informatics and Natural Sciences, Department of Earth Sciences,  Meteorological Institute, Bundesstra\ss{}e 55,  20146 Hamburg, Germany}


\author[0000-0003-4507-9384]{Abbie Donaldson} 
\affiliation{Institute for Astronomy, University of Edinburgh, Royal Observatory, Edinburgh, EH9 3HJ, UK}

\author[0000-0002-9925-0426]{Daniel Gardener}  
\affiliation{Institute for Astronomy, University of Edinburgh, Royal Observatory, Edinburgh, EH9 3HJ, UK}

\author{Dennis Crake} 
\affiliation{Institute for Astronomy, University of Edinburgh, Royal Observatory, Edinburgh, EH9 3HJ, UK}

\author[0000-0002-0167-3595]{Sedighe Sajadian}
\affiliation{Department of Physics,  Isfahan University of Technology,  Isfahan 84156-83111, Iran}

\author{Valerio Bozza}  
\affiliation{Dipartimento di Fisica ``E.R. Caianiello'', Universit{\`a} di Salerno,  Via Giovanni Paolo II 132, 84084 Fisciano, Italy}
\affiliation{Istituto Nazionale di Fisica Nucleare,  Sezione di Napoli, Strada Comunale Cinthia, 80126 Napoli, Italy}

\author{Jesper Skottfelt}  
\affiliation{Centre for Electronic Imaging,  School of Physical  Sciences, The Open University, Milton Keynes, MK7 6AA, UK}

\author[0000-0002-3202-0343]{Martin Dominik} 
\affiliation{University of St Andrews, Centre for Exoplanet Science, SUPA School of Physics \& Astronomy, North Haugh, St Andrews, KY16 9SS, UK}

\author{J. Fynbo}
\affiliation{Niels Bohr Institute, University of Copenhagen, Blegdamsvej 17, 2100 Copenhagen, Denmark}

\author[0000-0001-8870-3146]{Tobias C. Hinse} 
\affiliation{University of Southern Denmark, Department of Physics, Chemistry and Pharmacy, Campusvej 55, 5230 Odense M, Denmark}

\author{Markus Hundertmark} 
\affiliation{Astronomisches Rechen-Institut,  Zentrum f{\"u}r Astronomie der Universit{\"a}t Heidelberg  (ZAH), 69120 Heidelberg, Germany}

\author[0000-0002-7084-5725]{Sohrab Rahvar} 
\affiliation{Department of Physics, Sharif University of Technology,  PO Box 11155-9161 Tehran, Iran}

\author[0000-0002-3807-3198]{John Southworth} 
\affiliation{Astrophysics Group, Keele  University, Staffordshire, ST5 5BG, UK}

\author{Jeremy Tregloan-Reed}
\affiliation{Instituto de Astronomia y Ciencias Planetarias de Atacama,  Universidad de Atacama, Copayapu 485,  Copiapo, Chile}

\author{Mike Kretlow}
\affiliation{Instituto de Astrof\'{i}sica de Andaluc\'{i}a, Granada, Spain}

\author{Paolo Rota}
\affiliation{Dipartimento di Fisica ``E.R. Caianiello'', Universit{\`a} di Salerno,  Via Giovanni Paolo II 132, 84084 Fisciano, Italy}
\affiliation{Istituto Nazionale di Fisica Nucleare, Sezione di Napoli, Strada Comunale Cinthia, 80126 Napoli, Italy}

\author[0000-0002-6830-476X]{Nuno Peixinho}
\affiliation{Instituto de Astrof\'{i}sica e C\^{e}cias do Espa\c{c}o, Departamento de F\'{i}sica, Universidade de Coimbra, PT3040-004 Coimbra, Portugal}

\author{Michael Andersen}  
\affiliation{Niels Bohr Institute, University of Copenhagen, Blegdamsvej 17, 2100 Copenhagen, Denmark}


\author{Flavia Amadio} 
\affiliation{Centre for ExoLife Sciences, Niels Bohr Institute, University of Copenhagen, {\O}ster Voldgade 5, 1350 Copenhagen,  Denmark}

\author{Daniela Barrios-L\'opez}
\affiliation{Centro de Astronom{\'{\i}}a, Universidad de Antofagasta, Av.\ Angamos 601, Antofagasta, Chile}

\author{{Nora Soledad} {Castillo Baeza}}
\affiliation{Centro de Astronom{\'{\i}}a, Universidad de Antofagasta, Av.\ Angamos 601, Antofagasta, Chile}




\begin{abstract}
The NASA's Double-Asteroid Redirection Test (DART) was a unique planetary defence and technology test mission, the first of its kind.  The main spacecraft of the DART mission impacted the target asteroid Dimorphos, a small moon orbiting asteroid (65803) Didymos, on 2022 September 26. The impact brought up a mass of ejecta which, together with the direct momentum transfer from the collision, caused an orbital period change of $33\pm 1$ minutes, as measured by ground-based observations. We report here the outcome of the optical monitoring campaign of the Didymos system from the Danish $1.54\,\rm{m}$ telescope at La Silla around the time of impact. The observations contributed to the determination of the changes in the orbital parameters of the Didymos-Dimorphos system, as reported by \citet{Thomas2023}, but in this paper we focus on the ejecta produced by the DART impact. We present photometric measurements from which we remove the contribution from the Didymos-Dimorphos system using a H-G photometric model. Using two photometric apertures we determine the fading rate of the ejecta to be $0.115\pm0.003\,\magd$ (in a $2\arcsec$ aperture) and $0.086\pm0.003\,\magd$ ($5\arcsec$) over the first week post-impact. After about 8 days post-impact we note the fading slows down to $0.057\pm0.003\,\magd$ ($2\arcsec$ aperture) and $0.068\pm0.002\,\magd$ ($5\arcsec$). We include deep-stacked images of the system to illustrate the ejecta evolution during the first 18 days, noting the emergence of dust tails formed from ejecta pushed in the anti-solar direction, and measuring the extent of the particles ejected sunward to be at least 4000 km.
\end{abstract}

\keywords{Lightcurves (918) -- 
Asteroid satellites (2207) -- 
Apollo asteroids (58) -- 
Planetary probes (1252) 
}


\section{Introduction} \label{sec:intro}

The Double-Asteroid Redirection Test (DART) mission was a world-first practical planetary defence test \citep{Rivkin2021}. The DART spacecraft was a $580\,\rm{kg}$ probe that crashed into asteroid Dimorphos, a small moon of asteroid Didymos, on  2022 September 26 at 23:14:24.183 UTC \citep{Daly2023}. 
One of the principal aims of the DART mission was to change the orbital period of Dimorphos around Didymos in a measurable way, in order to understand and quantify the effect of collisions on Near-Earth-Objects, with the long-term aim of getting a deeper understanding about humanity’s possibility to protect Earth against potential future asteroid impact. This goal was successfully achieved \citep{Thomas2023,Cheng2023}. 
On top of the primary goals, the mission also provided a number of unique opportunities to study other science questions related to asteroids. For example, it provided a rare insight into the mechanics of asteroid collisions that happen naturally in the Solar System, but for which we have mostly indirect evidence in the dynamical structure of the main asteroid belt \citep[more specifically the asteroid families, e.g. reviews by][]{Nesvorny2015,Michel2015} and the rubble-pile appearance of asteroids \citep[discussed extensively, for example by][]{Walsh2018,Michel2020}. Direct observations of the dust ejected by the impact, which we will refer to as an ejecta cloud, post impact can also be used to investigate active asteroids, objects on typically asteroidal orbits displaying activity. The possible activity scenarios for this type of object include outgassing, mass shedding and recent impacts \citep[e.g.][and references therein]{Jewitt2015}. While the features of sublimation-driven activity are well studied in context of comets, asteroid impacts present transient events that are impossible to predict and thus monitor early on. The DART impact was therefore a rare opportunity to collect observations of the formation of an ejecta cloud and tail virtually from the moment of collision. 
 The immediate impact aftermath was observed by the Light Italian Cubesat for Imaging of Asteroids (LICIACube) that accompanied the DART spacecraft \citep{Dotto2023,LICIACube} and further measurements were made at various telescopes, for example by the Hubble Space Telescope \citep{Li2023,Jewitt2023}, the 8m Very Large Telescope in Chile \citep{Opitom2023, Murphy2023}, an array of smaller facilities \citep{Lin2023, kareta, Lister2023, Moskovitz2023}, and by amateur astronomers \citep{Graykowski2023}. We report on our observations obtained as a part of this global campaign.  In Section~\ref{sec:observations} we summarise the data collected in our photometric monitoring campaign with the Danish $1.54~\rm{m}$ telescope, as part of the MiNDSTEp consortium that operates the  telescope for 6 months each year. In Section~\ref{sec:lcanalysis} we present results of light-curve analysis, reporting the dimming rate of the ejecta. Finally, in Section~\ref{sec:morphology} we discuss the morphology of the ejecta cloud and tail that formed after the impact, as imaged through the first 18 days after the impact.

\section{Observations} %
\label{sec:observations}

\begin{deluxetable*}{cccc cccc ccccc}
\tablenum{1} 
\tablecaption{List of the lightcurves of asteroid (65803) Didymos presented in this study. %
	\label{tab:obs}}             
\tablewidth{0pt}
\tablehead{
  {UT Date}                     & {JD}& {Len.} & {Exp.}& {Seeing} & {$R_{\odot}$} 	&	 {$\Delta$} 	&	 {$\alpha$} & {$V_{mag}$} 	&	 {$\lambda_o$} 	&	  $\beta_o$     		 	\\	
  {[yyyy-mm-dd] [hh:mm:ss.sss]} &     & {[h]} &{[s]}	 & {["]} &	 {[AU]}  	&	   {[AU]}   	&	 {[$\degr$]}  & {[mag]}	&	 {[$\degr$]} 	&	 {[$\degr$]}    	
}
\startdata
2022-Sep-16	08:28.8	&	2459838.881	&	1.5	&	10	&	1.9	&	1.078	&	0.096	&	38.9	&	14.696	&	5.2	&	-41	\\	
2022-Sep-20	10:37.5	&	2459842.757	&	7.1	&	20	&	3	&	1.065	&	0.087	&	43.7	&	14.585	&	13.8	&	-44.8	\\															
2022-Sep-25	34:11.2	&	2459847.774	&	6.2	&	7	&	2.6	&	1.051	&	0.078	&	50.7	&	14.508	&	29.3	&	-48.9	\\	
2022-Sep-26	22:05.4	&	2459848.682	&	1	&	7	&	2	&	1.048	&	0.077	&	52.1	&	14.511	&	32.7	&	-49.4	\\	\hline 
2022-Sep-27	26:22.5	&	2459849.768	&	6.7	&	7	&	1.6	&	1.045	&	0.075	&	53.7	&	14.492	&	36.9	&	-49.8	\\	
2022-Sep-28	56:46.6	&	2459850.706	&	4.7	&	7	&	1.9	&	1.043	&	0.074	&	55.1	&	14.498	&	40.7	&	-50.1	\\	
2022-Sep-29	14:03.9	&	2459851.76	&	6.8	&	7	&	1.6	&	1.04	&	0.073	&	56.7	&	14.507	&	45.1	&	-50.2	\\	
2022-Sep-30	09:33.7	&	2459852.757	&	7.2	&	7	&	1.8	&	1.038	&	0.073	&	58.2	&	14.545	&	49.3	&	-50.2	\\	
2022-Oct-01	46:07.1	&	2459853.824	&	3.8	&	6	&	1.5	&	1.036	&	0.072	&	59.7	&	14.556	&	53.9	&	-49.9	\\	
2022-Oct-02	05:12.0	&	2459854.754	&	4.9	&	6	&	1.3	&	1.034	&	0.072	&	61.1	&	14.592	&	57.8	&	-49.6	\\	
2022-Oct-03	57:55.7	&	2459855.707	&	1	&	6	&	1.5	&	1.032	&	0.071	&	62.5	&	14.598	&	61.8	&	-49	\\	
2022-Oct-04	09:00.9	&	2459856.673	&	1.2	&	6	&	2.1	&	1.03	&	0.071	&	63.8	&	14.634	&	65.8	&	-48.3	\\	
2022-Oct-05	46:03.0	&	2459857.699	&	2.2	&	12	&	2.8	&	1.028	&	0.071	&	65.2	&	14.671	&	69.8	&	-47.4	\\	
2022-Oct-06	13:05.3	&	2459858.801	&	4.5	&	10	&	2.8	&	1.026	&	0.071	&	66.6	&	14.710	&	73.9	&	-46.3	\\	
2022-Oct-07	19:18.5	&	2459859.763	&	4.3	&	8	&	1.6	&	1.024	&	0.072	&	67.8	&	14.773	&	77.4	&	-45.2	\\	
2022-Oct-08	45:27.6	&	2459860.782	&	5.5	&	7	&	2.5	&	1.023	&	0.072	&	69	&	14.807	&	80.8	&	-44	\\	
2022-Oct-09	00:16.0	&	2459861.834	&	2.6	&	9	&	1.9	&	1.021	&	0.073	&	70.1	&	14.868	&	84.1	&	-42.6	\\	
2022-Oct-10	55:51.5	&	2459862.789	&	4.8	&	8	&	2.4	&	1.02	&	0.074	&	71.1	&	14.926	&	86.9	&	-41.3	\\	
2022-Oct-11	56:57.4	&	2459863.706	&	1.1	&	7	&	2.4	&	1.019	&	0.075	&	71.9	&	14.981	&	89.4	&	-40.1	\\	
2022-Oct-12	26:12.7	&	2459864.852	&	1.2	&	10	&	1.7	&	1.018	&	0.076	&	72.8	&	15.037	&	92.3	&	-38.4	\\	
2022-Oct-13	25:17.0	&	2459865.851	&	1.1	&	8	&	1.8	&	1.017	&	0.077	&	73.6	&	15.088	&	94.7	&	-37	\\	
2022-Oct-14	44:41.3	&	2459866.823	&	0.8	&	8	&	1.7	&	1.016	&	0.078	&	74.2	&	15.134	&	96.8	&	-35.6	\\	
\enddata

\tablecomments{
The table includes details of each lightcurve collected in our Didymos observing campaign. All lightcurves were collected in Bessel R filter. 
The columns list the observing circumstances for each night: 
 Universal Time (UT) `{Date}'  and Julian Date (`{JD}'){at} the middle of the exposure series taken on a given night, 
    the total length (`{Len.}') (in hours) of the observing sequence,
    the exposure time (`{Exp.}') (in seconds) of the individual images in the sequence,
    the approximate `{Seeing}' (in arcseconds), estimated based on the FWHM of selected background stars, 
	the heliocentric (`$r_{\odot}$') and geocentric (`$\Delta$') distances measured in AU, 
    the solar phase angle (`$\alpha$'), 
    the apparent `$V_{mag}$' brightness as predicted with the  H=18.16, G=0.2 model, \citep{Pravec2012},
	the {observer-centred} ecliptic longitude (`$\lambda_o$') and latitude (`$\beta_o$').
	Each row represents a single-night lightcurve which is divided into multiple segments if it spans more than about an hour. The horizontal line splits pre-impact and post-impact data sets.}
\end{deluxetable*}

We observed the Didymos-Dimorphos system with the Danish $1.54\,\rm m $ telescope (DK), located at La Silla observatory in Chile (MPC code W74). The Danish Faint Object Spectrograph and Camera (DFOSC) instrument, with field of view $13\farcm5\times13\farcm5$, was used in imaging mode. A series of images in Bessel R filter were taken on each observing night, as listed in Table~\ref{tab:obs}. 
The object was moving rapidly across the field, but we decided to use sidereal tracking and short exposures to keep both the asteroid and the background stars from trailing.
In order to maintain consistency of relative photometry and monitor the ejecta features, the field was regularly adjusted to keep the asteroid within the central area of the detector while maximising the overlap between background stars in any sequence of observations. For the purpose of performing relative photometry the lightcurves were divided into individual `segments'. The main practical difference between the segments is the selection of background stars available. The lightcurve segments were usually around 1-hour long, but the exact time they covered depended on how often the observers changed the field.

\subsection{Lightcurve extraction}
\label{sec:lcextraction}

\begin{deluxetable*}{ccc ccc ccc c}
\tablenum{2}
\tablecaption{Outputs of the photometric monitoring of the asteroid (65803) Didymos with the DK $1.54\,\rm{m}$ telescope. %
	\label{tab:res}}             
\tablewidth{0pt}
\tablehead{
  {UT Date}    & {Mid-seq. time}&  {$H(1,1,\alpha)_{2"}$} & {$rms_{2"}$} 	&{$H(1,1,\alpha)_{5"}$} & {$rms_{5"}$} 	&	 {$\Delta H_{2"}$}  	&	 {$\Delta H_{5"}$} & {{$l_\odot$}}  	&	 {$l_{-\odot}$} 	     		 	\\	
  {[yyyy-mm-dd]} &  {[d] since impact}   & {[mag]}	 & {[mag]} &	 {[mag]}  	&	   {[mag]}   	&	 {[mag]} 	&	 {[mag]} 	&	 {[km]} & {[km]}   	
}
\startdata
2022-Sep-16	&	-10.58717	&	19.703	&	0.051	&	\mdash	&	\mdash	&	0.081	&	\mdash	&	\mdash	&	\mdash	\\
2022-Sep-20	&	-6.71068	&	19.919	&	0.050	&	19.849	&	0.044	&	0.168	&	0.098	&	\mdash	&	\mdash	\\
2022-Sep-25	&	-1.69432	&	20.064	&	0.044	&	20.073	&	0.039	&	0.124	&	0.133	&	\mdash	&	\mdash	\\ 
2022-Sep-26	&	-0.78605	&	20.127	&	0.043	&	20.118	&	0.048	&	0.150	&	0.141	&	\mdash	&	\mdash	\\ \hline
2022-Sep-27	&	0.30026	&	18.961	&	0.027	&	18.865	&	0.023	&	-1.060	&	-1.157	&	3200	&	3500	\\
2022-Sep-28	&	1.23803	&	19.021	&	0.013	&	18.949	&	0.014	&	-1.039	&	-1.111	&	2500	&	6900	\\
2022-Sep-29	&	2.29171	&	19.193	&	0.021	&	19.100	&	0.018	&	-0.912	&	-1.006	&	4200	&		\\
2022-Sep-30	&	3.28858	&	19.365	&	0.029	&	19.236	&	0.022	&	-0.782	&	-0.912	&	3800	&	14900	\\
2022-Oct-01	&	4.35563	&	19.530	&	0.021	&	19.374	&	0.019	&	-0.663	&	-0.819	&	3300	&	13500	\\
2022-Oct-02	&	5.28555	&	19.659	&	0.025	&	19.469	&	0.023	&	-0.573	&	-0.764	&	3800	&	15400	\\
2022-Oct-03	&	6.23883	&	19.794	&	0.031	&	19.582	&	0.025	&	-0.479	&	-0.692	&	2100	&	6500	\\
2022-Oct-04	&	7.20487	&	19.893	&	0.033	&	19.664	&	0.024	&	-0.420	&	-0.649	&	2100	&	5700	\\
2022-Oct-05	&	8.23059	&	19.906	&	0.051	&	19.662	&	0.037	&	-0.449	&	-0.693	&	1900	&	7800	\\
2022-Oct-06	&	9.33270	&	20.086	&	0.043	&	19.817	&	0.031	&	-0.312	&	-0.581	&	2200	&	14500	\\
2022-Oct-07	&	10.29535	&	20.228	&	0.053	&	19.940	&	0.034	&	-0.206	&	-0.494	&	2500	&	10900	\\
2022-Oct-08	&	11.31351	&	20.321	&	0.037	&	20.050	&	0.029	&	-0.150	&	-0.421	&	1900	&	11700	\\
2022-Oct-09	&	12.36546	&	20.394	&	0.028	&	20.134	&	0.023	&	-0.112	&	-0.372	&	2200	&	9400	\\
2022-Oct-10	&	13.32073	&	20.515	&	0.055	&	20.256	&	0.040	&	-0.022	&	-0.281	&	1500	&	11200	\\
2022-Oct-11	&	14.23816	&	20.586	&	0.038	&	20.350	&	0.033	&	0.021	&	-0.214	&	\mdash	&	7400	\\
2022-Oct-12	&	15.38348	&	20.679	&	0.049	&	20.454	&	0.056	&	0.085	&	-0.141	&	\mdash	&	\mdash	\\
2022-Oct-13	&	16.38283	&	20.734	&	0.040	&	20.523	&	0.032	&	0.115	&	-0.096	&	1300	&	\mdash	\\
2022-Oct-14	&	17.35464	&	20.798	&	0.037	&	20.607	&	0.031	&	0.158	&	-0.033	&	\mdash	&	\mdash	\\
\enddata

\tablecomments{
The first column is the UT date at which the observations were taken, for easy linking with Table~\ref{tab:obs}. Second column, labelled `Mid-seq. time', shows time (in days) since the DART mission impact (which happened at JD 2459849.46806) at the middle of the observing sequence. The columns labelled $H(1,1,\alpha)$ include the Johnson-V mean magnitude for an observing sequence reduced to $1\,\rm{au}$ heliocentric and geocentric distance, measured in $2\arcsec$ and $5\arcsec$-radius apertures. The $rms$ column illustrates the scatter of the measurements going into the mean brightness, which comes from both measurement uncertainty and shape effects. The $\Delta H$ columns include the excess brightness (as compared with the  H=18.16, G=0.2 model, \citet{Pravec2012}). The final two columns illustrate the estimated lower limit on the extent of the post-impact dust cloud in the Sunward ($l_\odot$) and antisolar ($l_{-\odot}$) directions. The dust cloud extents were measured using the \noisechisel{} low-surface-brightness feature extraction tool. Note that the listed dust cloud extents are generally underestimated in the anti-solar direction, as explained in detail in the main text.}
\end{deluxetable*}

\begin{figure}

\gridline{
\trimfig{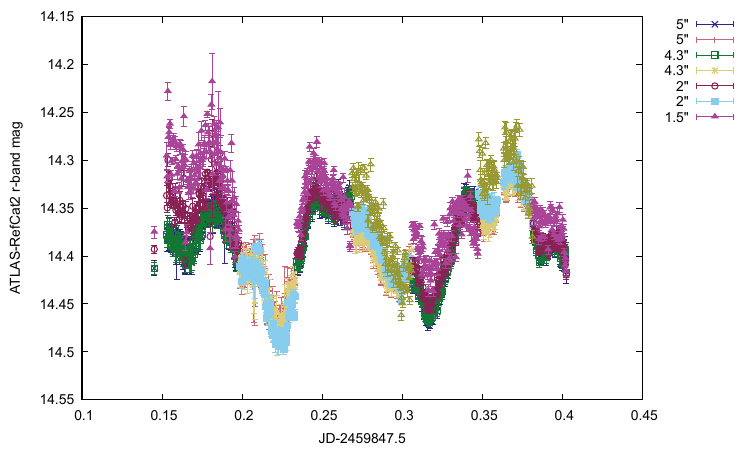}{0.48\textwidth}{0 0 1.5cm 0}{(a) 25 September 2022}
\trimfig{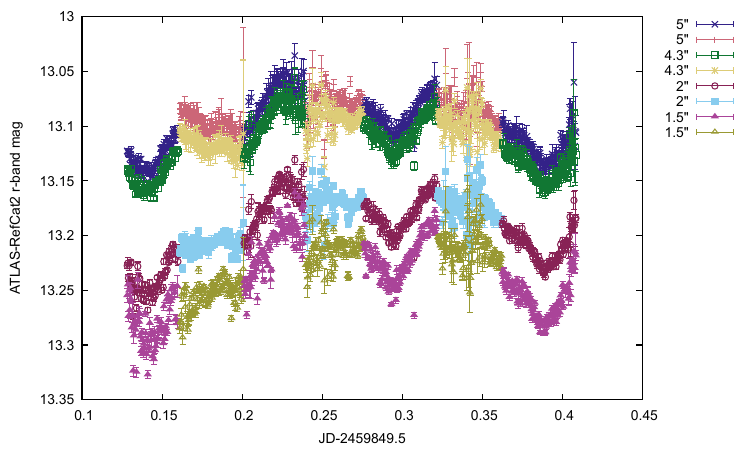}{0.52\textwidth}{0.5cm 0 0 0}{(b) 27 September 2022}
}

\caption{A sample of multi-aperture light curves collected at the Danish telescope, calibrated to ATLAS-RefCat2 r-band. Panel (a) shows an example pre-impact light curve (taken with  the four fixed-angular-size apertures discussed in the main text), and panel (b) shows an example of multiaperture post-impact light curve. The horizontal axis scale is in days since 0~UT of a given observing night. Each night is divided into `segments' corresponding to different background fields used for relative light-curve extraction. Each segment and aperture size is marked with a different set of colours and symbols. Full light-curve data set is presented in Figs.~\ref{fig:preimpact:all} and~\ref{fig:postimpact:all}. \label{fig:LC:example}}
\end{figure}

The data reduction and photometry was performed using custom python procedures utilising astropy \citep{astropy:2013,astropy:2018,astropy:2022}. Imaging frames were reduced using standard methods in CCD imaging: bias subtraction and flat-field correction using twilight flats. 
The asteroid brightness was measured relative to the background stars in each image. Between 5 and 10 stars were selected fulfilling two criteria: they were not as bright as to be overexposed, and have a catalogue colour $(g-r)<1.5$ (or $<1.2$ whenever sufficient number of stars fulfilling this criterion were available). The stellar and asteroid brightness was measured using aperture photometry. Due to large amount of ejecta observed in the weeks after the impact, and to ensure consistency with other data sets used for the photometric study of the system \citep{Thomas2023}, we decided to use a few fixed-on-sky-size apertures, namely: $1\farcs5$, $2\arcsec$, $4\farcs3$, and $5\arcsec$. 
We note that due to the seeing conditions throughout our campaign (Table~\ref{tab:lc}) there is little difference between the measurements made in the $1\farcs5$ and $2\arcsec$ apertures. Similarly, the photometry in $4\farcs3$ and $5\arcsec$ gives comparable results, as both include considerable contribution from the sunlight reflected by the dispersing ejecta cloud. However there is a noticeable difference in the photometric behaviour of the system between the small apertures, at the order of seeing, and photometry in larger apertures, so in the further discussion below we will focus on the $2\arcsec$ and $5\arcsec$ radius apertures.
In Table~\ref{tab:lc} (available on-line) we present the individual lightcurve segments calibrated to the ATLAS-RefCat2 r-band magnitudes with the calviacat python package \citep{calviacat, Tonry2018} using the methods outlined by \citet{Donaldson2022}. For the purpose of extracting a relative lightcurve we used the same set of comparison stars as for the calibration. While computing the calibrated magnitude for the asteroid system the colour $(g-r)=0.52$ was adopted \citep{Pravec2022}. 

\subsection{Morphology extraction}

To illustrate the changing dust environment of the Didymos-Dimorphos system we have used the \gnuastro{} software suite \citep{gnuastro,noisechisel_segment_2019} to create deep-stacked images for each hour on the first night post-impact and for each night in the rest of the sequence. We used the \noisechisel{} routine to remove background sky from images before stacking and then cropped to create $501\times501$ pixel images on the first night and $701\times1501$ images for the entire sequence, keeping Didymos always in the same place in the new frames. The images were then combined by taking sigma-clipped mean value for each pixel. This helped remove the contribution from background stars trailing across the faint ejecta features.  
The images were then processed with \noisechisel{}. This subroutine was designed for detection of low-surface-brightness features in analysis of extragalactic observations. We tested it here to be used as a method of `activity' assessment for the Didymos-Dimorphos system. The subroutine carves out signal close to background level, by comparing the statistical properties of pixel values in pixels adjacent to an initial high signal-to-noise-ratio detection with statistical properties of the detection-free background. We use this method to investigate the shape of dust cloud surrounding the asteroid system post-impact.

\section{Lightcurve analysis}
\label{sec:lcanalysis}

The contribution of ejecta to the post-impact light curve is immediately apparent. Before the impact, the asteroid brightness measurements were independent of the aperture size within the photometric error bars. 
From the night of impact there is a clear offset between the brightness measured in different apertures (Fig.~\ref{fig:LC:example}). The extent of the ejecta cloud and the eventual emergence of the ejecta tail made it impractical to attempt setting up an aperture that would encompass all of the ejecta. The light curve thus includes the effects due to shape and rotation of Didymos, mutual eclipses between Didymos and Dimorphos, and ejecta. 
Short-term variations (i.e. over the course of a few hours in a single night) are dominated by the first two elements, and the ejecta contribution can be approximated as a constant to allow lightcurve decomposition methods to be applied \citep[e.g.,][]{Pravec2006}. The results of this analysis were used to measure the change in Dimorphos' orbital period and are described elsewhere \citep{Thomas2023,Scheirich2023}. 

\begin{figure*}
\gridline{
\fig{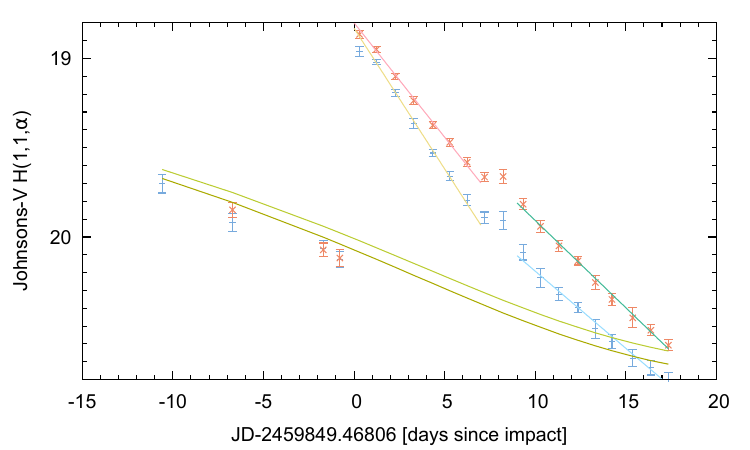}{0.5\textwidth}{(a)} \fig{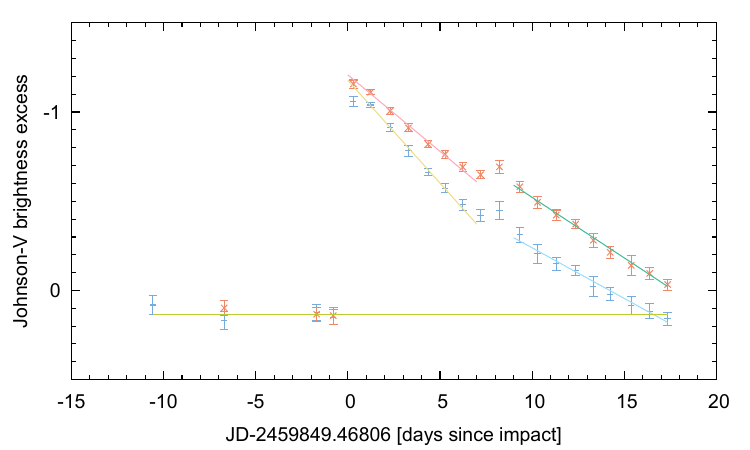}{0.5\textwidth}{(b)}
}

\caption{The figure shows the apparent fading of the Didymos-Dimorphos system in the Johnson-V band. The blue crosses represent the averaged brightness in the $2\arcsec$ aperture for each night and red `x'-symbols correspond to the brighness in the $5\arcsec$ aperture. Panel (a) shows the heliocentric and geocentric-distance-corrected brightness of the asteroids and ejecta cloud; the straight coloured lines represent linear trend fits to subsets of data; the curved lines show the asteroid brightness as predicted by the photometric models.  In panel (b) we remove the contribution from the Didymos-Dimorphos system by using the photometric model from \citet{Pravec2012} and re-measure the brightness fading rates - illustrated again with coloured straight lines; the horizontal line illustrates the straight-line fit to the pre-impact photometry.
		\label{fig:LC:stat}}
\end{figure*}

In order to investigate the long-term photometric behaviour of the ejecta we have taken the average brightness from each night, translated it to Johnson-V system by approximating the ATLAS-RefCat-r with PanSTARRS-r and using the correction
$(V-r) = 0.248$,
where $(V - r)$ is the difference between Johnson-V magnitude and our calibrated r-magnitude. The correction is estimated using $(g-r)=0.52$ colour for Didymos \citep{Pravec2022}, and a relevant formula from \citet{Tonry2012}. The conversion to Johnsons-V magnitude was needed to facilitate comparison with photometric models that are usually expressed in this system \citep[e.g.][]{Pravec2012}. We then 
reduce the V-magnitude to $1\,\rm{au}$ distance from the Sun and Earth, the $H(1,1,\alpha)$ magnitude. 
The $H(1,1,\alpha)$ magnitude measurements made in the $2\arcsec$-radius aperture as well as in the $5\arcsec$-radius aperture are summarised in Table~\ref{tab:res}. Between the pre-impact observations and the night of impact the brightness increased by about 1.3 mag in the $5\arcsec$ aperture, and by 1.2 mag in the $2\arcsec$ aperture relative to what was expected from pre-impact photometric models and measured shortly before the impact (Fig.~\ref{fig:LC:stat}). The  brightness then faded at $0.105\pm0.002\,\magd$ in the $5\arcsec$ aperture, and at a slightly quicker rate of  $0.133\pm0.001\,\magd$  in the $2\arcsec$ aperture, until about 8 days post-impact. At that time the system brightness jumped by about 0.2 mag above the expected trend in all apertures, but continued fading after the event. Interestingly, the fading rate measured between day 9 and 18 is quicker in the $5\arcsec$ aperture ($0.120\pm0.002\,\magd$) than in the $2\arcsec$ aperture ($0.109\pm0.002\,\magd$). 

The brightness fading is a combination of several factors:
the solar phase angle at which the system is observed is increasing, with the asteroid moving away from Earth the size (measured in km) of the aperture is getting smaller, and the amount of detectable ejecta in each aperture is varying due to the ejecta movement.
The variation of the ejecta flow and brightness is of the highest interest in the context of the DART impact event. However, in order to make this kind of measurement, it is necessary to first subtract the non-ejecta-related brightness variations from the observed values. To this purpose
we include the two (pre-impact) photometric models, described with the absolute magnitude at zero phase angle (H) and slope parameter (G), as the two curves in the lower part of
Fig.~\ref{fig:LC:stat} (a). The upper of these two lines represents the H=18.12, G=0.15 model that is used to provide brightness estimates in the Horizons ephemerides system\footnote{\url{https://ssd.jpl.nasa.gov/horizons/}}, 
while the lower of the two lines is the 
 H=18.16, G=0.2 model from \citet{Pravec2012}.  We note that both the literature model and the solution adapted in the Horizons ephemerides system overestimate the brightness of the Didymos-Dimorphos system before impact, the measured system brightness lay below both lines. This is likely due to the unusually large phase angles at which the observations were collected. 
 The data that were used for the determination of the  H=18.16, G=0.2 model covered phase angles $2-40\deg$ \citep{Kitazato2004,Pravec2006} while in the time span of our observations the solar phase angle changes from about $50\degr$ around the time of impact to $75\degr$ at the end of the sequence. 
 The~magnitudes from the model by \citet{Pravec2012} are in better agreement with our observed pre-impact magnitudes of the system than the alternative Horizons-model (while both models agree with our determined slope in the pre-impact magnitudes). We therefore adopt the \citet{Pravec2012} model to deduce
 the brightness excess due to ejecta alone (Fig.~\ref{fig:LC:stat} (b)). Measurement on the pre-impact lightcurves shows that the asteroid brightness was overestimated by the model by about $0.132\pm0.018$ mag. The fading trends are shallower when measured on just the brightness excess alone, and also change around the 8-day mark. The fading slows down, from initially being $0.115\pm0.003\,\magd$ (in $2''$ aperture) and $0.086\pm0.003\,\magd$ (in $5''$ aperture) over the first week to $0.057\pm0.003\,\magd$ ($2''$) and $0.068\pm0.002\,\magd$ ($5''$). This fading rate convolves the rate of ejecta escaping the system with solar-phase-angle effects that are different for different ejecta size regimes \citep{Lolachi2023}. 
  In the phase-angle regime covered by our observations the brightness of very fine sub-mm material should appear to slightly increase, while for larger (mm to cm) material it should drop, and drop even faster for larger boulders (meter-sized) and for the asteroids themselves.
For a typical cometary coma phase function \citep{Schleicher}, which is dominated by micron-scale grains, the same reflecting area of dust would appear $\sim17\%$ brighter due to the changing phase angle observed between the impact and mid-October. Over the same phase angle range, the brightness of the larger material (according to the H-G photometric model) reduces by $\sim43\%$.  

Hubble and MUSE observations are consistent with fine-grained material leaving the system quickest, as expected, as it is pushed away by the solar radiation, with larger particles lingering longer \citep{Li2023,Opitom2023}. We estimate the acceleration due to solar radiation pressure using
\begin{equation}
    a_{srp} = \frac{3 L_\odot}{16 \pi c R_\odot^2 \rho r}Q_{srp},
\end{equation}

where the density of the dust grains $\rho$ is assumed to be $3000\,\textrm{kg}~\textrm{m}^{-3}$ for S-type asteroid material, the coefficient $Q_{srp} \approx 1$ for grains larger than the dominant wavelengths of sunlight, and we take the solar distance $R_\odot$ as a constant (at its average value of 1.03 au) over the time frame considered here. We find that grains of radius $r = 1 \mu \textrm{m}$ are accelerated to a projected distance of $> 5\arcsec$ (more than 340 km along the anti-Sun direction) by the time of the first post-impact observations, assuming that they had zero initial velocity relative to Didymos (so, in an abstract scenario of a dust cloud co-moving with the system). This means that these would have already left the photometric apertures used, and therefore the brightening of small grains at increasing phase angle can be ignored in explaining the overall brightness evolution. The picture is  complicated by the fact that the ejecta clearly had non-zero initial velocities, and that material ejected into the cone oriented opposite DART's impact velocity, which was approximately sun-ward (see Sec.~\ref{sec:morphology}), will first be decelerated before passing through the photometric aperture on its way into the tail direction. This is a proposed explanation for the peak in brightness 8 days after impact \citep{Li2023}, but a significant amount of micron-sized grains would have had to have an initial sunward velocity of $\sim 350$ m~s$^{-1}$ to be accelerated back past the asteroid at +8 days from impact, and would have reached a projected $> 15\arcmin$ in the sunward direction before turning back, filling the DFOSC field of view, which doesn't match the observed extent of the dust. This is size dependent: 10 micron grains could explain a +8 day bump with initial velocities and maximum sunward extents 10 times lower, which are a better match to the observed morphology. 

In principle, one could calculate the total brightness within an aperture at a given time by summing the contribution over different grain size bins:
\begin{equation}
    f(t) = \sum_r N(r) \zeta(r,t) \pi r^2 p_V \Phi(r,\alpha),
\end{equation}
where $N(r)$ is the initial population of grains of a given size, $\zeta(r,t)$ describes the fraction remaining in the aperture at time $t$, $p_V$ is the single scattering albedo of the grains (assumed to be the same for all grains) and $\Phi(r,\alpha)$ is the size dependent phase integral. For an assumed initial size distribution of ejecta with a continuous power-law form $N(r) \propto r^{-q}$ this relationship should be integrable. In practice there are too many unknowns within $\zeta$ (primarily the 3D distribution of initial velocities, and any size-dependence on those) for this to be of much use in fitting the fading relationship seen in Fig.~\ref{fig:LC:stat}. Modelling the full motion of ejecta is beyond the scope of this work, but has been tackled by others \citep[e.g.,][]{Moreno2023}.

\section{Morphology}
\label{sec:morphology}

\begin{figure*}

\gridline{
\trimfig{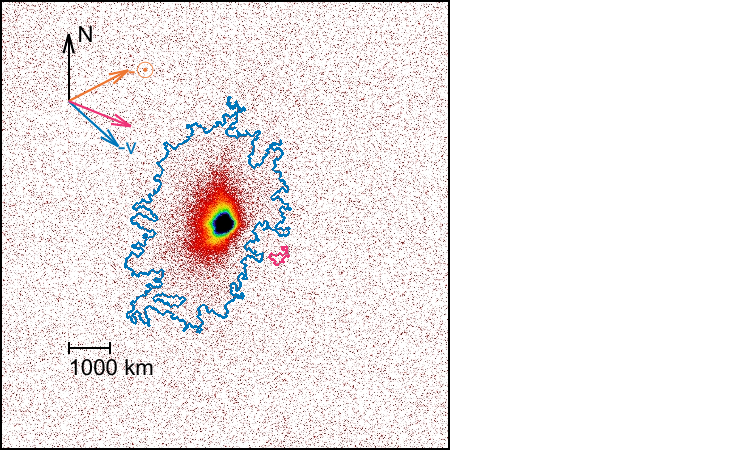}{0.25\textwidth}{0 0 5.05cm 0}{(a) 03:04:32.807 - 03:50:13.575}
\trimfig{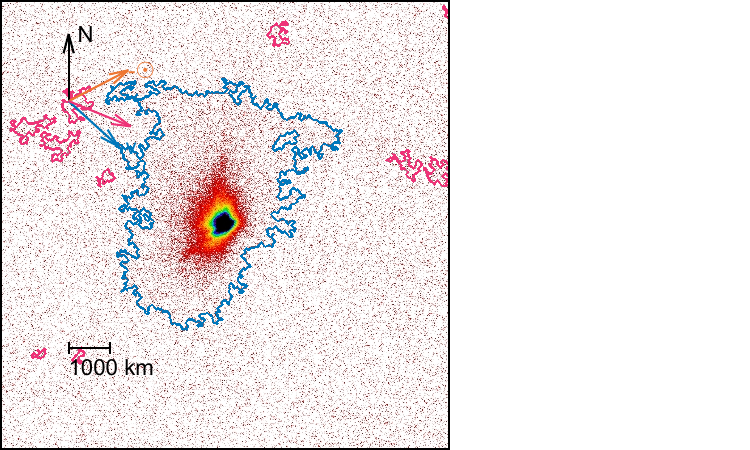}{0.25\textwidth}{0 0 5.05cm 0}{(b) 03:51:59.420 - 04:49:00.847}
\trimfig{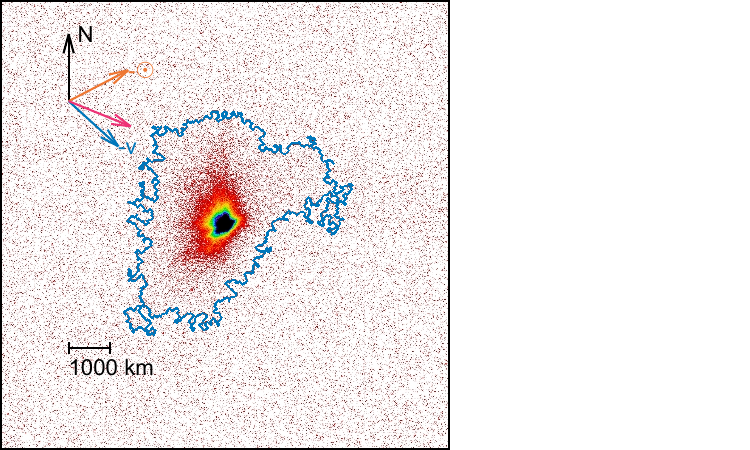}{0.25\textwidth}{0 0 5.05cm 0}{(c) 04:49:38.319 - 05:44:20.656}
\trimfig{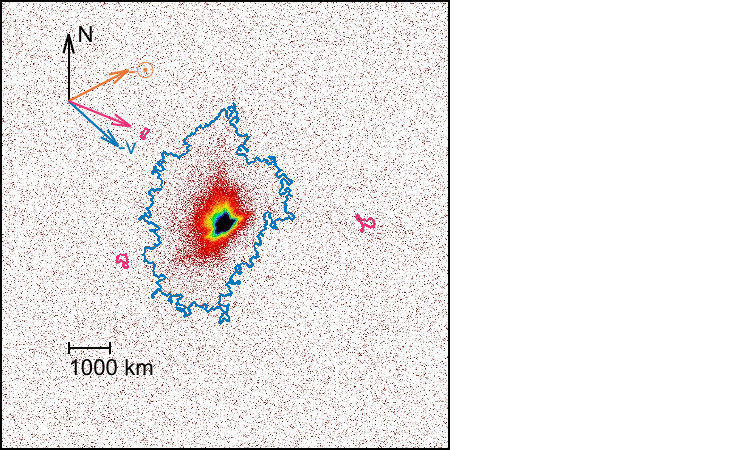}{0.25\textwidth}{0 0 5.05cm 0}{(d) 05:45:12.698 - 06:36:30.803}
}

\gridline{
\trimfig{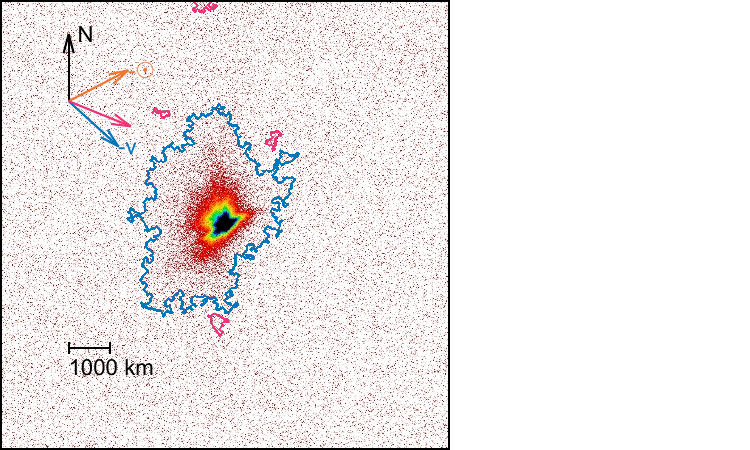}{0.25\textwidth}{0 0 5.05cm 0}{(e) 06:37:27.336 - 07:42:55.663}
\trimfig{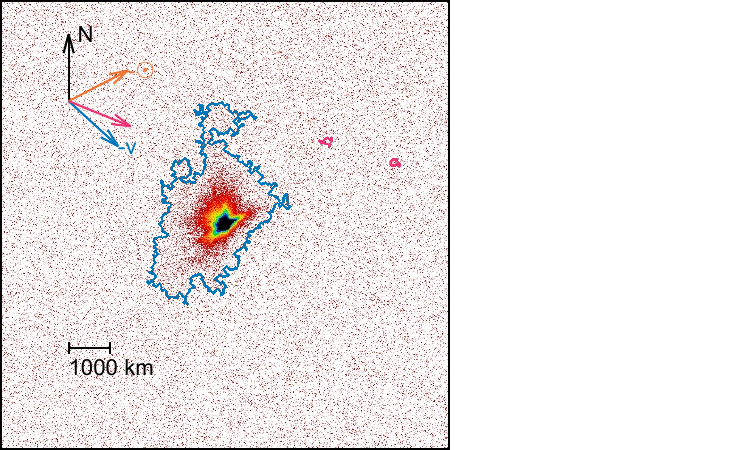}{0.25\textwidth}{0 0 5.05cm 0}{(f) 07:43:54.735 - 08:41:01.763}
\trimfig{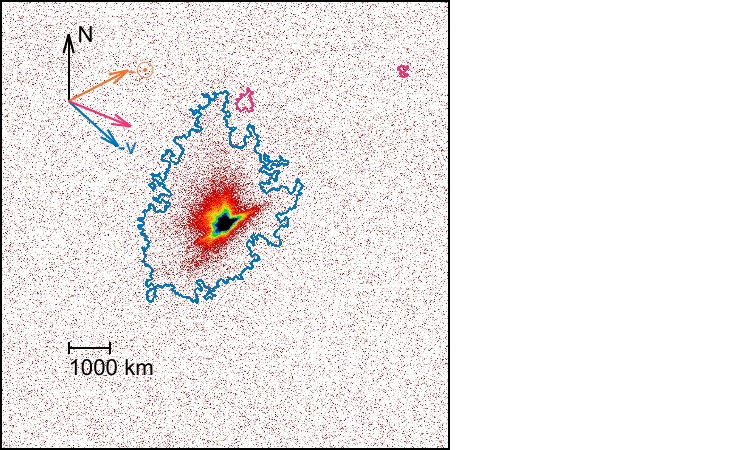}{0.25\textwidth}{0 0 5.05cm 0}{(g) 08:41:53.566 - 09:48:05.157}
}

\caption{Ejecta evolution in the DK images over the first observing night post-impact.
The dominant feature in these images is the ejecta cone, fanning out in the direction the DART probe arrived from. The tail extending in the anti-solar direction can be seen forming in panels d-g.
Each panel represents a stack of images with the exposure-start UT-time on 2022 September 27 for the first and last frame in each sequence included in the panel captions, individual images were exposed for 7s. 
Each image stack was produced by aligning images on Didymos, taking a  $501\times501$ pixel cutout and then taking a sigma-clipped mean of corresponding pixels in each image. The colour-scale goes from white and red for low values to dark blue and black for high values; the minimum and maximum mapped pixel values are taken to be $0.5\sigma$ below and $25\sigma$ above the mean pixel value in each stack. The \noisechisel{} detection contours are overlain on the images; the blue outline marks the specific detection that includes the asteroid plus the ejecta, and magenta outlines all other detections. The arrow set in the corner of the image represent north (black arrow labelled `$N$', east is to the left), the anti-solar direction (orange, `$-\odot$'), the direction opposite to the system's projected orbital velocity (blue,`$-v$'), and the direction from which the DART probe approached the system (unlabelled magenta arrow). 
Image scale is indicated with a ruler corresponding to $1000\,\rm{km}$ at the asteroid system.
		\label{fig:ejecta:firstnight}}
\end{figure*}

\begin{figure*}

\gridline{
\trimfig{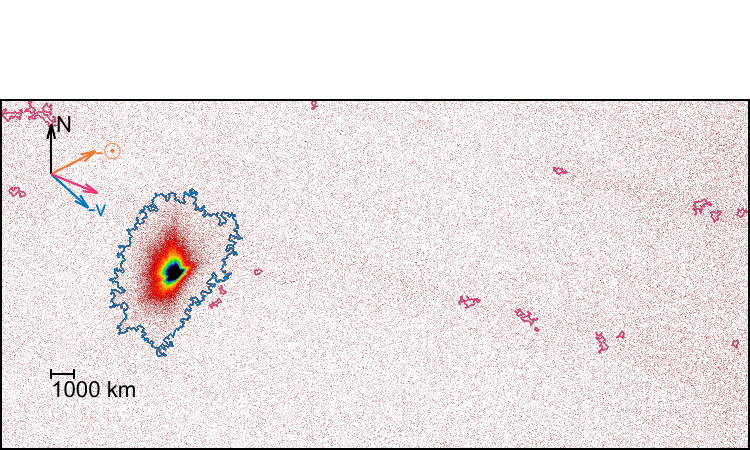}{0.33\textwidth}{0 0 0 1.65cm}{(a) 2022 September 27}
\trimfig{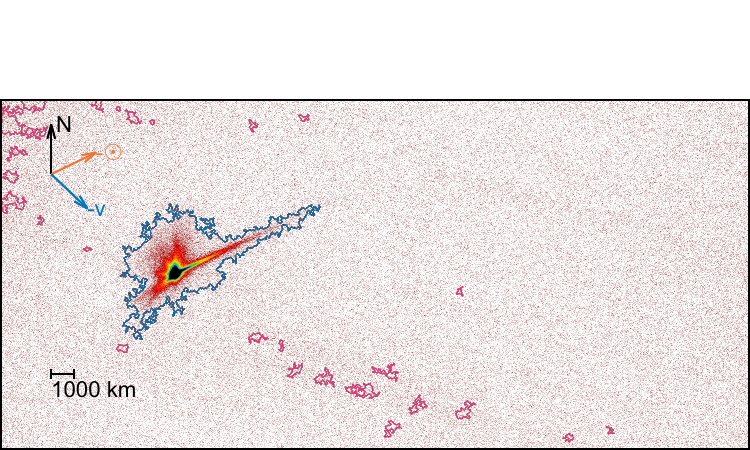}{0.33\textwidth}{0 0 0 1.65cm}{(b) 2022 September 28}
\trimfig{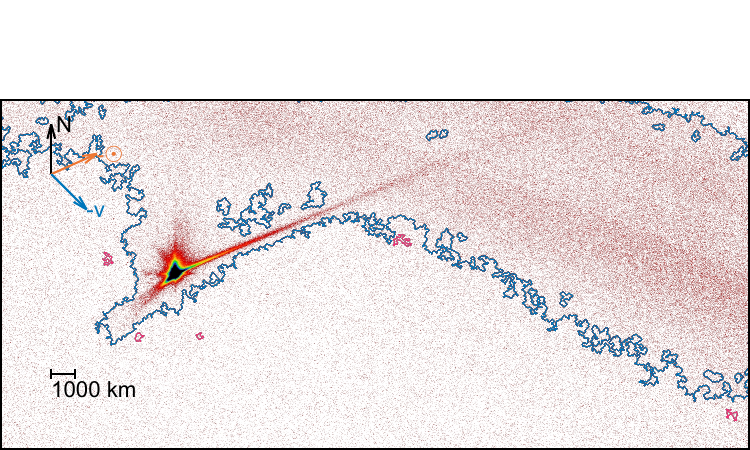}{0.33\textwidth}{0 0 0 1.65cm}{(c) 2022 September 29}
}

\gridline{
\trimfig{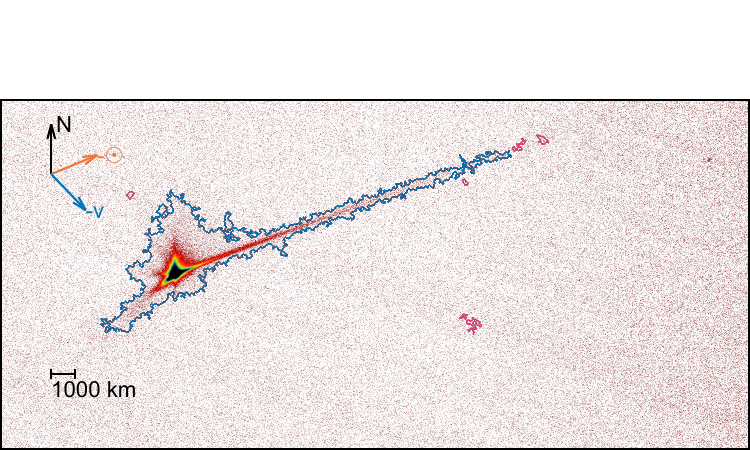}{0.33\textwidth}{0 0 0 1.65cm}{(d) 2022 September 30}
\trimfig{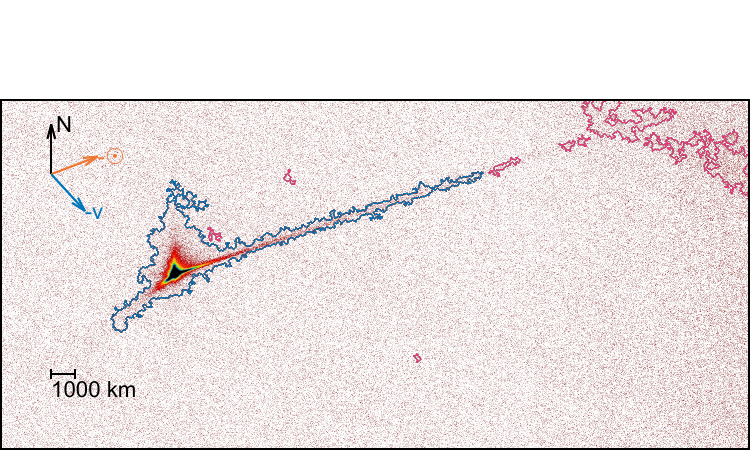}{0.33\textwidth}{0 0 0 1.65cm}{(e) 2022 October 1}
\trimfig{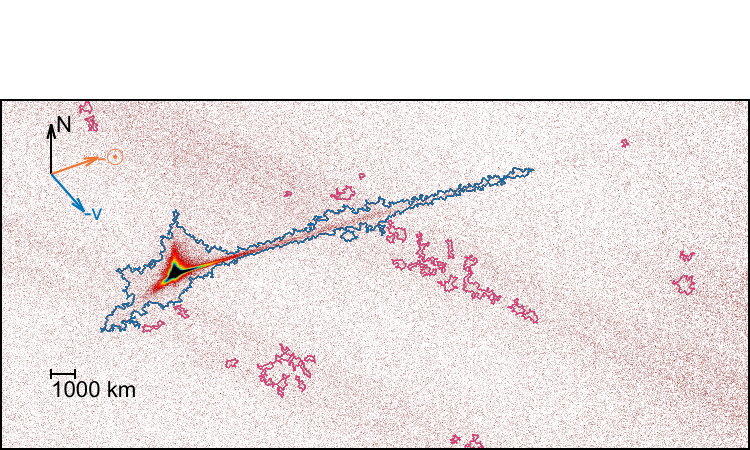}{0.33\textwidth}{0 0 0 1.65cm}{(f) 2022 October 2}
}

\gridline{
\trimfig{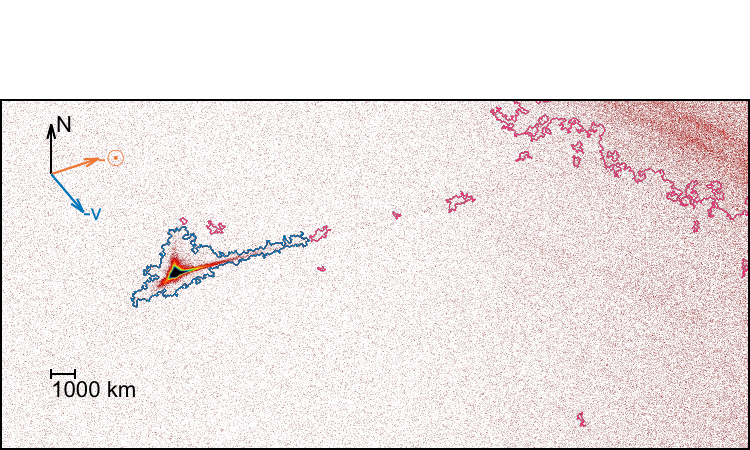}{0.33\textwidth}{0 0 0 1.65cm}{(g) 2022 October 3}
\trimfig{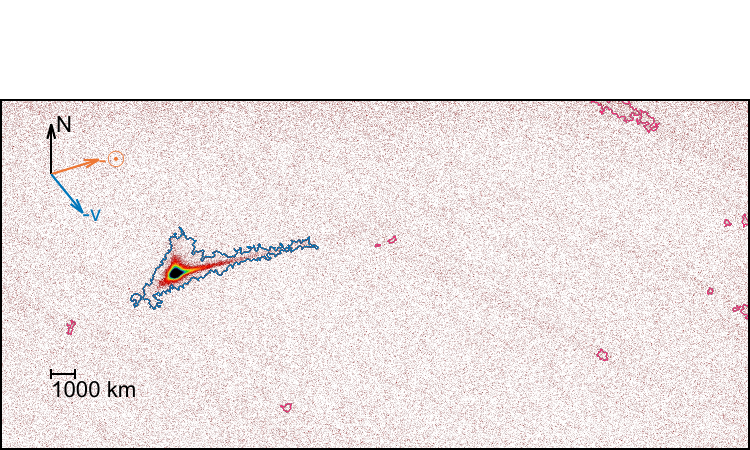}{0.33\textwidth}{0 0 0 1.65cm}{(h) 2022 October 4}
\trimfig{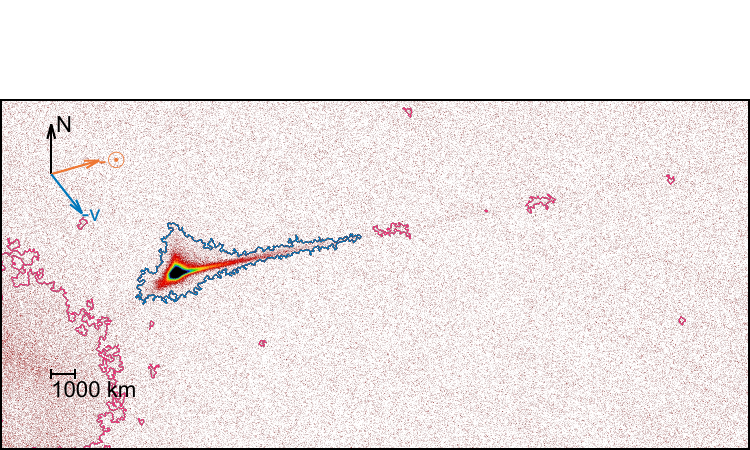}{0.33\textwidth}{0 0 0 1.65cm}{(i) 2022 October 5}
}

\caption{Illustration of the ejecta cloud and tail evolution throughout our post-impact observing campaign.
The tail extending in the anti-solar direction (indicated with the orange arrow), can seen forming in panels (a) and is clearly identifiable through the rest of our campaign. 
Each frame is a $1501 \times 701$ cutout, created by stacking all the images taken throughout each night in the intervals listed in Table~\ref{tab:obs}. 
The captions for individual panels denote the UT date the observations were taken. The stacking was done through aligning all images taken on a given night and then taking a sigma-clipped mean of corresponding pixels. More details in caption for Figure~\ref{fig:ejecta:firstnight}.
		\label{fig:ejecta:allnights}}
\end{figure*}

\addtocounter{figure}{-1}

\begin{figure*}
\gridline{
\trimfig{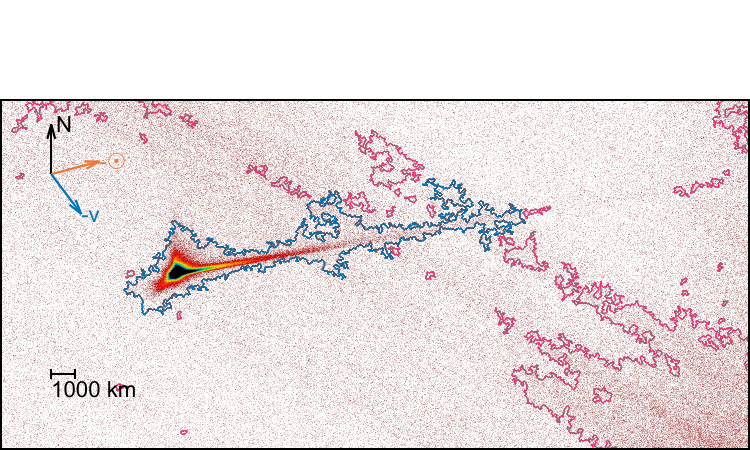}{0.33\textwidth}{0 0 0 1.65cm}{(j) 2022 October 6}
\trimfig{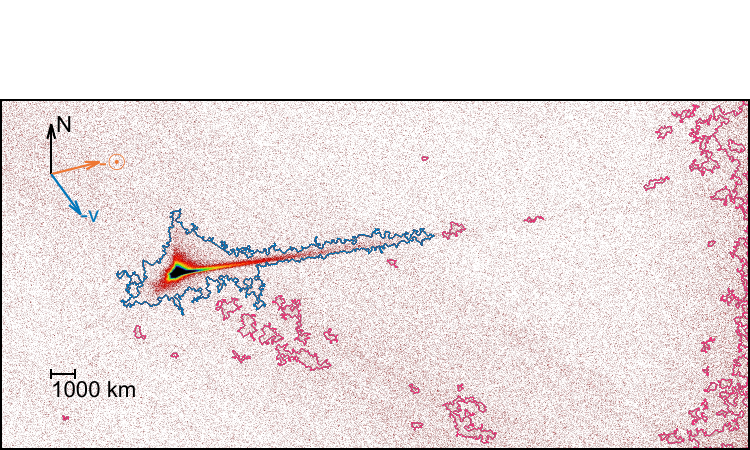}{0.33\textwidth}{0 0 0 1.65cm}{(k) 2022 October 7}
\trimfig{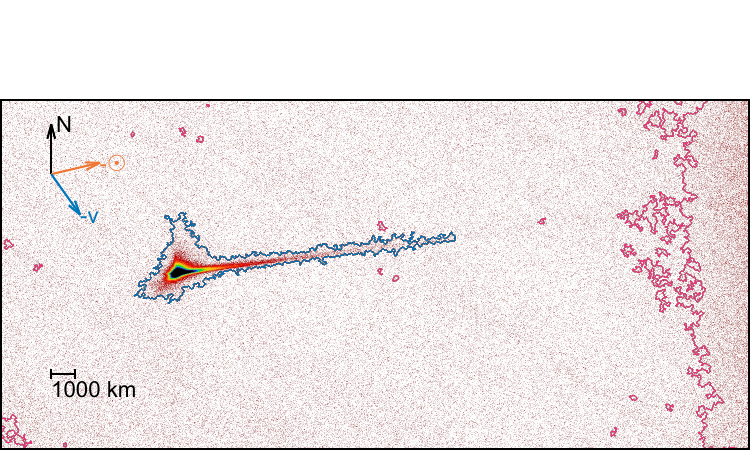}{0.33\textwidth}{0 0 0 1.65cm}{(l) 2022 October 8}
}

\gridline{
\trimfig{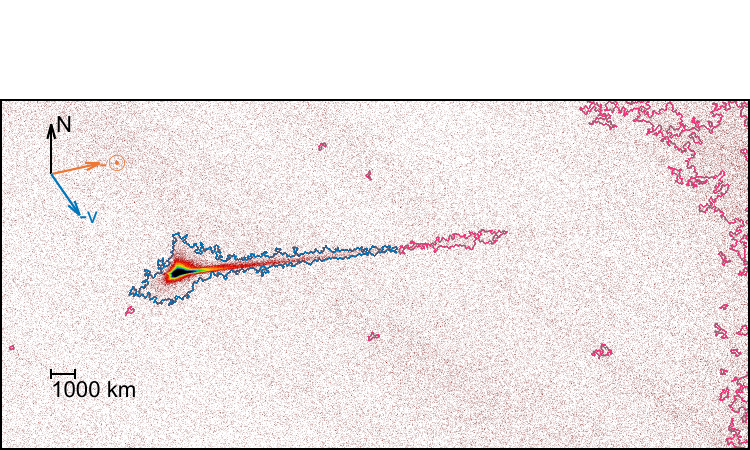}{0.33\textwidth}{0 0 0 1.65cm}{(m) 2022 October 9 }
\trimfig{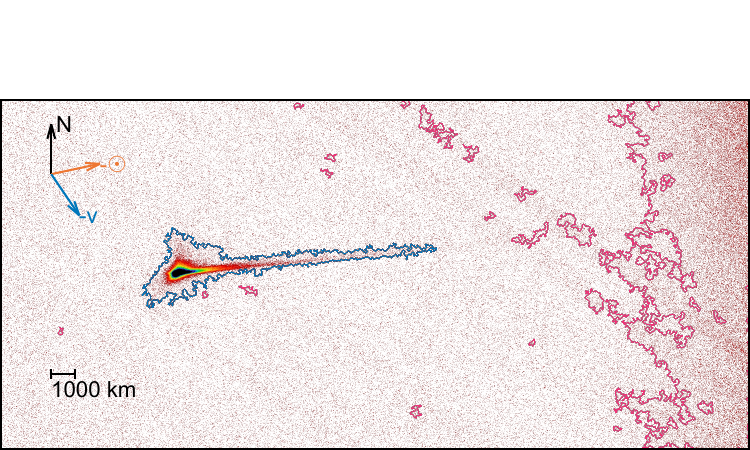}{0.33\textwidth}{0 0 0 1.65cm}{(n) 2022 October 10}
\trimfig{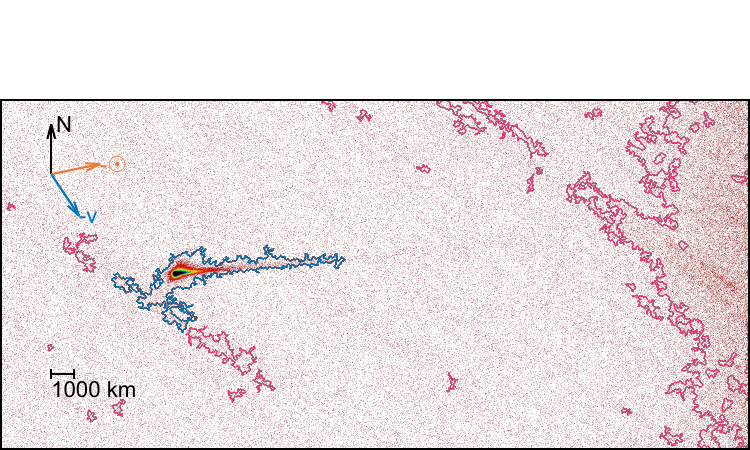}{0.33\textwidth}{0 0 0 1.65cm}{(o) 2022 October 11}
}

\gridline{
\trimfig{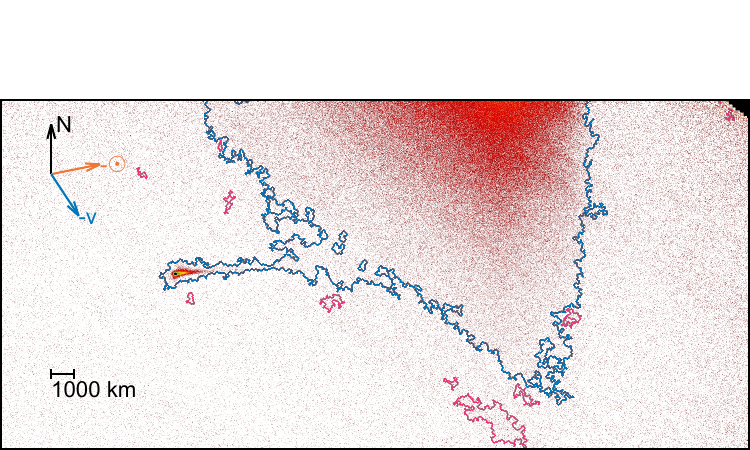}{0.33\textwidth}{0 0 0 1.65cm}{(p) 2022 October 12}
\trimfig{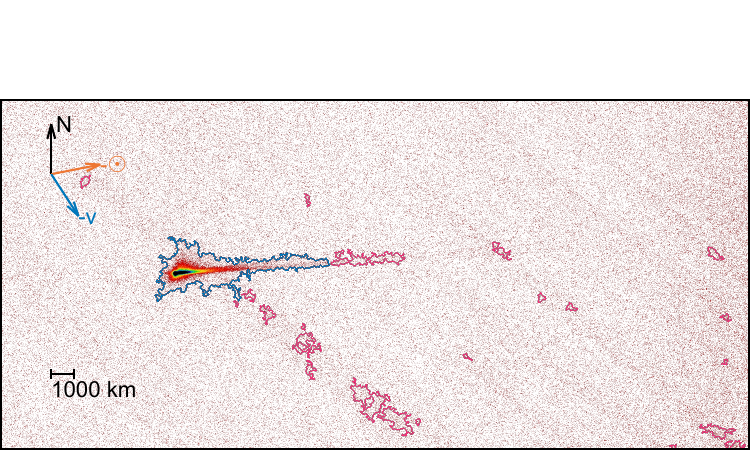}{0.33\textwidth}{0 0 0 1.65cm}{(q) 2022 October 13}
\trimfig{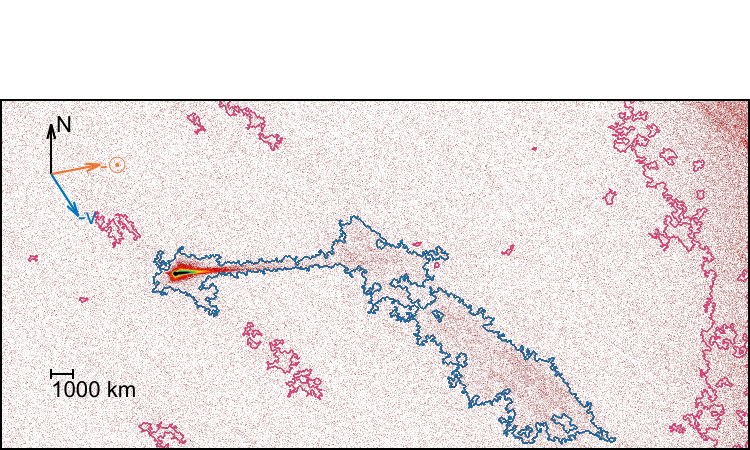}{0.33\textwidth}{0 0 0 1.65cm}{(r) 2022 October 14}
}
\caption{[Continued]}
\end{figure*}

Across the 18 days we  observed  post-impact the morphology of the ejecta cloud changed in dramatic ways. The detailed structure of the ejecta cloud changed across the seven hours that we  observed it on the first night, as illustrated on the seven panels in Fig.\ref{fig:ejecta:firstnight} (north is up in those images, and east to the left). Over the first night the dominant feature is the ejecta cone fanning out in the direction the DART probe arrived from \citep[at position angle $68\degr$ counted from North through East,][]{Cheng2023}.
The ejecta cone has a wide opening angle: one edge of the ejecta cone is projected almost exactly north in the images, while the other is south-east, extending almost Sun-ward.
The 18 panels in Fig.~\ref{fig:ejecta:allnights} correspond to all of the images collected on each night deep-stacked to highlight the ejecta features.
The sun-ward edge of the cone can be seen growing over the first few days post-impact and then gradually dissipating and shrinking as the solar wind pushes the material back. The northern edge of the ejecta cone also dissipates over the course of our observations as the material gets pushed approximately westward. 
The tail extending in the anti-solar direction, as marked by the direction of the orange arrow in Figs.~\ref{fig:ejecta:firstnight} and \ref{fig:ejecta:allnights}, that is clearly visible in later days can be identified by the end of the first post-impact night. 
There are also fragments of dust cloud moving away from Didymos and slowly dispersing. By the second night post-impact the ejecta tail is very clearly visible and spanning at least 320 pixels (about $7000\,\textrm{km}$).  Hubble observations from October 8th and 11th  clearly show a `double tail' feature \citep{Li2023}. We cannot resolve this feature in our observations that early, but it can be seen in images taken on October 13th and 14th. 

We used the \noisechisel{} procedure (from the \gnuastro{} suite) to establish an approximate outline of the ejecta cloud in the stacked images. The outline of detections containing the Didymos-Dimorphos system are marked with blue lines in Figs.~\ref{fig:ejecta:firstnight} and \ref{fig:ejecta:allnights}. We present approximate measurement of Sun-ward ($l_\odot$) and anti-solar ($l_{-\odot}$) extents of the ejecta cloud in Table~\ref{tab:res}. These were made using the contours of detections made on the $1501\times701$ image stacks, as shown in Fig.~\ref{fig:ejecta:allnights}. While useful for qualitative illustration of the cloud extent and particularly informative when it comes to the changing morphology of the ejecta cloud Sun-ward extent (for example Fig.~\ref{fig:ejecta:allnights}  panels (c), (d), (e)) even using this method we are not able to fully separate the faint anti-solar tail from the background noise; for example in the observations from 5 October (Fig.~\ref{fig:ejecta:allnights} (i)) the tail detection is broken into several disjointed detections. The brighter background stars also somewhat interfere with the detections (for example Fig.~\ref{fig:ejecta:allnights} (j) and (o)).

The maximum Sun-ward extent of the ejecta cloud estimated from this method is around $4000\,\rm{km}$, and there remains ejecta in this direction throughout the period of observation, suggesting that this is not dominated by small grains. Micron-sized grains would have cleared over the weeks of observation, or would have extended further towards the Sun if given sufficient initial velocity to still be present at least 8 days after impact, as discussed above. Grains of around 100 micron in size and initial Sun-ward velocities of $5-10\,\rm{m}\,\rm{s}^{-1}$ would have approximately the right extent and lifetime. However, there must be a faster moving component ($\sim100\,\rm{m}\,\rm{s}^{-1}$) within the ejecta, as the cloud is already more than $3000\,\rm{km}$ in extent in both Sun-ward and anti-Sun projected directions only 7 hours after the impact. The growth of the tail (already more than $6000\,\rm{km}$ in length by the second night) is consistent with the material at the end of it being of micron scale and accelerating under Solar radiation pressure. This material would quickly accelerate to distances greater than those measured over the following nights, but it is worth noting that the anti-Solar direction `tail' lengths are all minimum values, and the true tail rapidly extends beyond the field of view of DFOSC at a flux level below that which can be separated from the background with statistical significance. 

\section{Discussion and Conclusions}

The brightness decay rate we observe for the ejecta (Section~\ref{sec:lcanalysis}) agrees with observations from other observatories \citep{kareta,Graykowski2023} and is consistent with material slowly leaving the aperture due to acceleration by solar radiation pressure. While the fast growth of the tail we observe is consistent with it containing particles down to micron size (Section~\ref{sec:morphology}), Finson-Probstein modelling by \citet{Lin2023} suggests dominant ejecta grains to be considerably larger, in the mm to cm size range. These authors find an  expansion velocity in the anti-Sun (tail) direction of $31\,\rm{m}\,\rm{s}^{-1}$ on the 27th of September, consistent with acceleration of mm-cm particles by radiation pressure, but our deeper images (acquired approximately 14 hours earlier from Chile, versus \citealt{Lin2023}'s observations from Lulin observatory in Taiwan) already show a longer tail. More detailed Monte Carlo modelling of the ejecta motions by \citet{Moreno2023} fits the ejecta using a broken power law size distribution between micron and $5\,\rm{cm}$ radius particles. They find a model that contains two components (low and high velocity, ejected hemispherically and into an ejecta cone, respectively), with a third component needed to explain the appearance of the second tail. Both \citet{Lin2023} and \citet{Moreno2023} agree that this secondary dust release event is necessary to produce this tail and that the timing of this is consistent with the `bump' in the photometry we observe around 8 days after impact. 
Interestingly, an alternative model demonstrating the secondary tail to be an effect of changing viewing geometry of the ejecta cone created in the  initial impact event was recently presented \citep{Kim2023}, but it is inconsistent with the $\approx8$ day brightening we note (Section~\ref{sec:lcanalysis} and Fig.~\ref{fig:LC:stat}).
The initial velocities in the Monte Carlo model by \citet{Moreno2023} are considerably lower than those we find necessary, of order of $\rm{m}\,\rm{s}^{-1}$ even for the `fast' component. While these provide good fits to the long term evolution of the ejecta cone and tails seen in their imaging with small telescopes, such low velocities cannot explain the 1000s of km extent of the ejecta cloud we see in our deep stacked images on the first nights after impact (Section~\ref{sec:morphology} and Figure~\ref{fig:ejecta:allnights}). Observations \emph{in situ} from LICIACube \citep{Dotto2023,LICIACube} also favour higher ejecta speeds, of order 10s of $\rm{m}\,\rm{s}^{-1}$ for most of the clumps they see, and $300-500\,\rm{m}\,\rm{s}^{-1}$ for faster streamers, which would be consistent with the expansion we see.
Finally, it's worth noting that even though the photometric contribution from the ejecta fades to almost 0 over the short span of our observations \citep[and fades completely shortly after, in about 20 days;][]{Moskovitz2023,Lister2023}, the tail persists. These observations are therefore consistent with asteroid collisions being one of the mechanisms behind  observed main-belt-asteroid activity. It is a direct confirmation that while it would be difficult to observe a quickly dispersing ejecta cloud or brightness enhancement, which are only present for a few days after an impact, a dust tail can be observed long after the event. 

To summarise: 
\begin{itemize}
    \item We report observations of Didymos-Dimorphos asteroid system collected in September and  October 2022 with the Danish $1.54\,\rm{m}$ telescope at La Silla. The photometric measurements contributed to the already reported orbital period change detection \citep{Thomas2023}, but here we focused on long-term development of the system brightness and the ejecta morphology.
    \item Accounting for changing observing geometry the system exhibits excess brightness, compared to earlier photometric models, that comes from the ejecta cloud. The brightness of the ejecta cloud decreases post impact, increases shortly around 8 days after impact, and then continues to fade at a slower rate than before day 8. Those measurements are consistent with measurements at other observatories and with a secondary impact scenario. The dust tail forms in the first hours after impact and persists for weeks.     
    \item We used \gnuastro{} and \noisechisel{} to characterise morphological features of the dust cloud. We are able to assess minimum extents of the ejecta cloud in both Sun-ward and anti-solar directions, consistent with small dust grains ejected at 10s to 100s  of $\rm{m}\,\rm{s}^{-1}$, similar to velocities measured by LICIACube. 
\end{itemize}

\begin{acknowledgments}
We thank two anonymous referees for their kind comments which helped to improve the presentation of our results.
We thank all the staff at the ESO La Silla Observatory for their support. %
AR and CS acknowledge support from the UK Science and Technology Facilities Council. %
This project has received funding from the European Union’s Horizon 2020 research and innovation programme under grant agreement No 870403 (NEOROCKS).%
This work was supported by the DART mission, NASA Contract No. 80MSFC20D0004.
This research has received funding from the Europlanet 2024 Research Infrastructure (RI) programme. The Europlanet 2024 RI provides free access to the world’s largest collection of planetary simulation and analysis facilities, data services and tools, a ground-based observational network and programme of community support activities. Europlanet 2024 RI has received funding from the European Union’s Horizon 2020 research and innovation programme under grant agreement No. 871149.
NP acknowledges funding from Funda\c{c}\~ao para a Ci\^{e}ncia e a Tecnologia (FCT), Portugal, through the research grants UIDB/04434/2020 and UIDP/04434/2020.
PLP was partly funded by Programa de Iniciación en Investigación-Universidad de Antofagasta. INI-17-03.
UGJ, MA, 
and FA acknowledge support from the European Union H2020-MSCA-ITN-2019 under grant No. 860470 (CHAMELEON) and the Novo Nordisk Foundation Interdisciplinary Synergy Programme grant no. NNF19OC0057374.
This paper uses Paul Tol's Colour schemes and templates \url{https://personal.sron.nl/~pault/}.
For the purpose of open access, the author has applied a Creative Commons Attribution (CC BY) licence to any Author Accepted Manuscript version arising from this submission.
\end{acknowledgments}

\facility{Danish 1.54m Telescope}
\software{Astropy \citep{astropy:2013, astropy:2018, astropy:2022}, Gnuastro \citep{gnuastro,noisechisel_segment_2019}} 

\bibliography{Didymos_DK}{}
\bibliographystyle{aasjournal}

\appendix

\begin{deluxetable*}{ccccc}
\tablenum{A1}
\tablecaption{Optical lightcurves of the asteroid (65803) Didymos collected with the DK $1.54\,\rm{m}$ telescope. %
	\label{tab:lc}}             
\tablewidth{0pt}
\tablehead{
  {JD}    & {r}&  {$\Delta r$} &  	{Seg.} & {Ap.} 	 	     		 	\\	
  {[d]} &   {[mag]}	 & {[mag]} &	 {--}  	&	   {[\arcsec]}    	
}
\startdata
2459838.848829 & 14.571 & 0.004 & 01 & 1.5 \\
2459838.849275 & 14.597 & 0.005 & 01 & 1.5 \\
2459838.849729 & 14.570 & 0.003 & 01 & 1.5 \\
2459838.850193 & 14.556 & 0.004 & 01 & 1.5 \\
2459838.850637 & 14.556 & 0.004 & 01 & 1.5 \\
\enddata

\tablecomments{Table A1 is published in its entirety in the machine-readable format.
      A portion is shown here for guidance regarding its form and content.
      The columns are: 
      `JD' in days, the time at mid-exposure with no light-time correction;
      `r' in mag, magnitude callibrated to ATLAS-RefCat2 r-band;
      `$\Delta r$' in mag, the 1-sigma uncertainty in the magnitude;
      `Seg.', a numerical label indicating which `Segment' or constant-background-stars field in a given night the measurement belongs to;
      `Ap.' in $\arcsec$, the diameter of the aperture for the measurements. The table is sorted by the aperture size first and JD second.   }
\end{deluxetable*}

\begin{figure}
\gridline{\fig{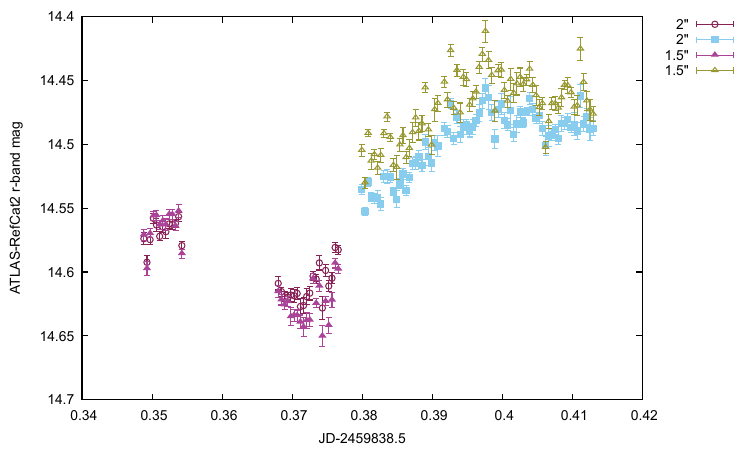}{0.5\textwidth}{(a) 16 September 2022}
\fig{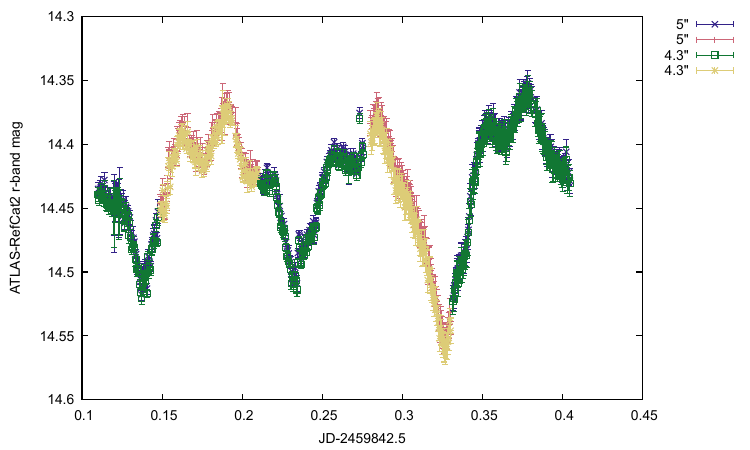}{0.5\textwidth}{(b) 20 September 2022}
}
\gridline{
\fig{65803_20220924_multiaperture_LC.pdf}{0.5\textwidth}{(c) 25 September 2022}
\fig{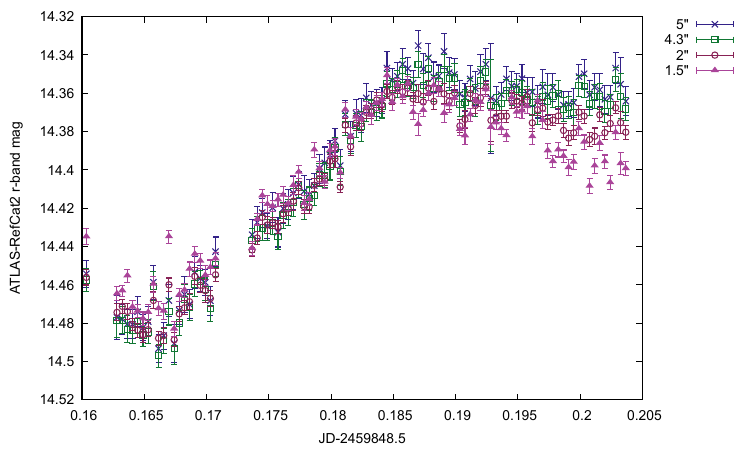}{0.5\textwidth}{(d) 26 September 2022}
}

\caption{Pre-impact lightcurves collected at the Danish telescope, calibrated to ATLAS-RefCat2 r-band. The horizontal axis scale is in days since 0~UT of a given observing night. Each night is divided into `segments' corresponding to different background fields used for relative lightcurve extraction. Each segment a is marked with a different colour and symbol. Due to low seeing and corresponding high noise in small apertures on 19th September plots of lightcurves in 1.5" and 2" apertures are omitted on that date. \label{fig:preimpact:all}}
\end{figure}

\begin{figure}
\gridline{\fig{65803_20220926_multiaperture_LC.pdf}{0.5\textwidth}{(a) 27 September 2022}
\fig{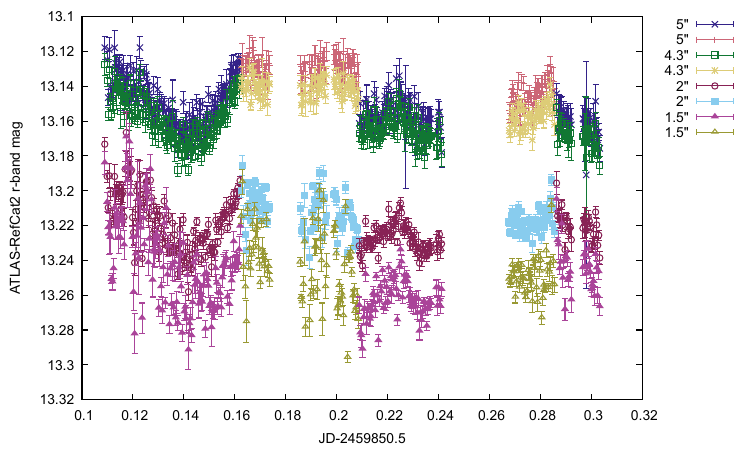}{0.5\textwidth}{(b) 28 September 2022}
}
\gridline{
\fig{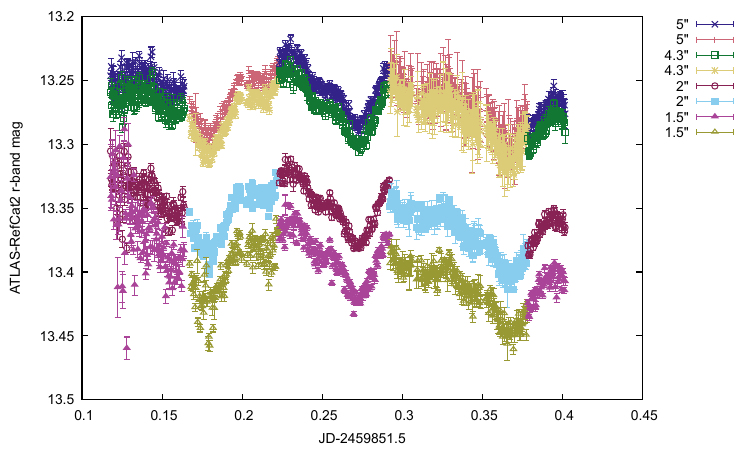}{0.5\textwidth}{(c) 29 September 2022}
\fig{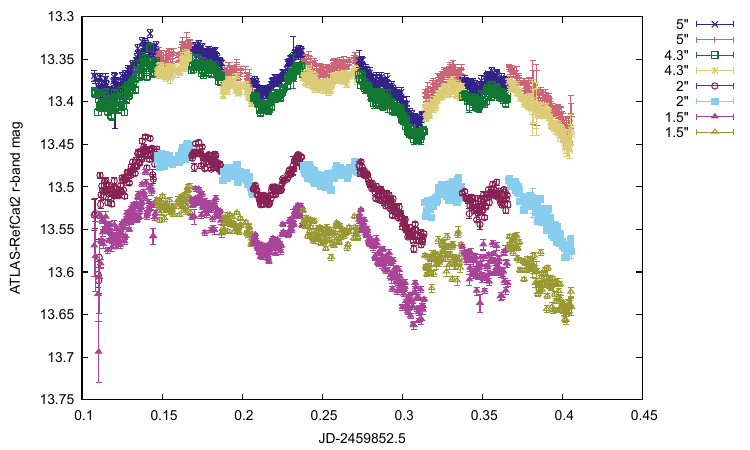}{0.5\textwidth}{(d) 30 September 2022}
}

\gridline{
\fig{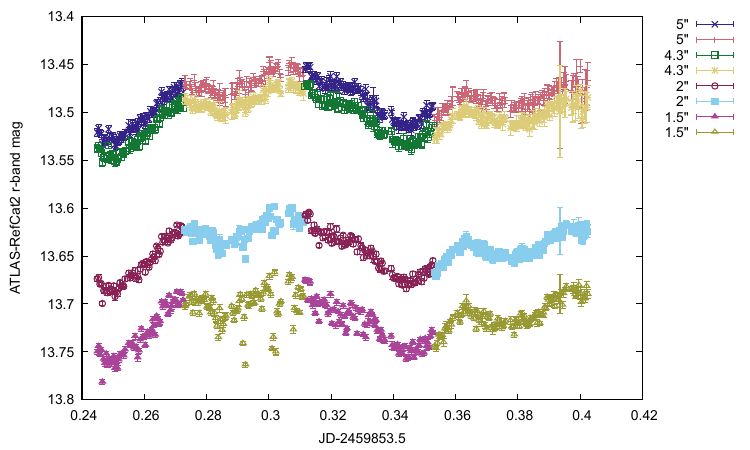}{0.5\textwidth}{(e) 1 October 2022}
\fig{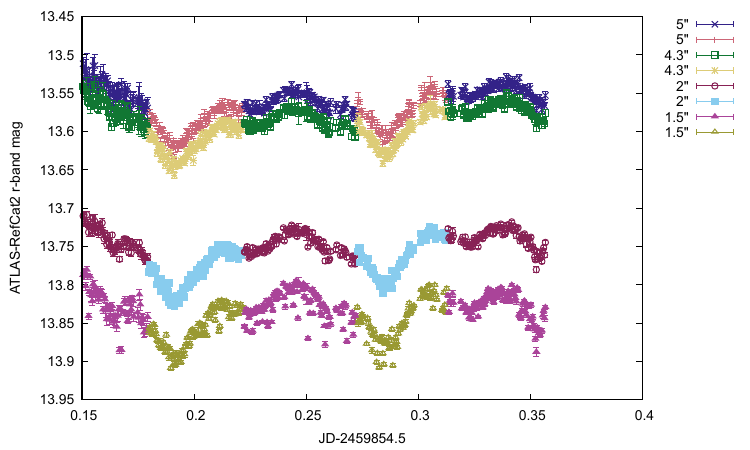}{0.5\textwidth}{(f) 2 October 2022}
}

\caption{Post-impact multi-aperture lightcurves collected at the Danish telescope, calibrated to ATLAS-RefCat2 r-band. The horizontal axis scale is in days since 0~UT of a given observing night. Each night is divided into `segments' corresponding to different background fields used for relative lightcurve extraction. Each segment and aperture size is marked with a different colour and symbol. \label{fig:postimpact:all}}
\end{figure}

\addtocounter{figure}{-1}
\begin{figure}

\gridline{
\fig{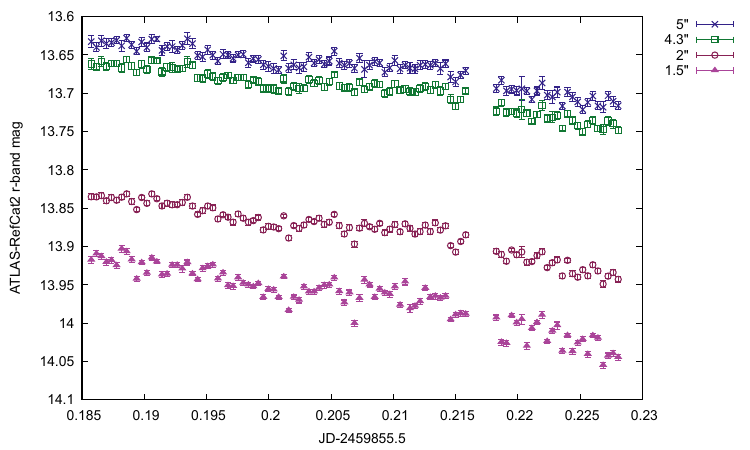}{0.5\textwidth}{(g) 3 October 2022}
\fig{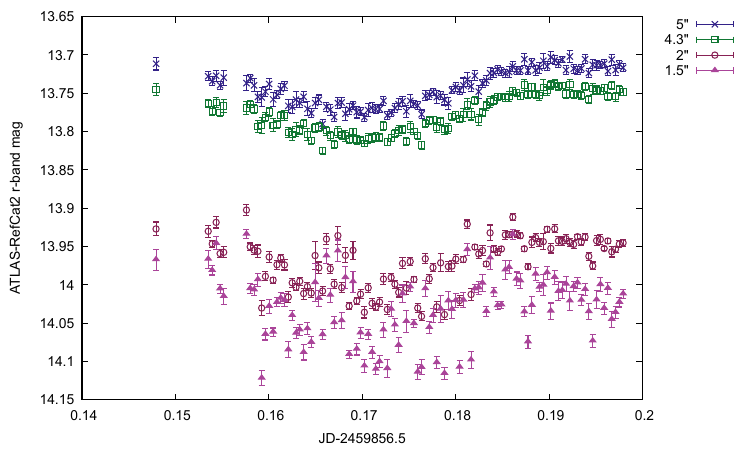}{0.5\textwidth}{(h) 4 October 2022}
}

\gridline{
\fig{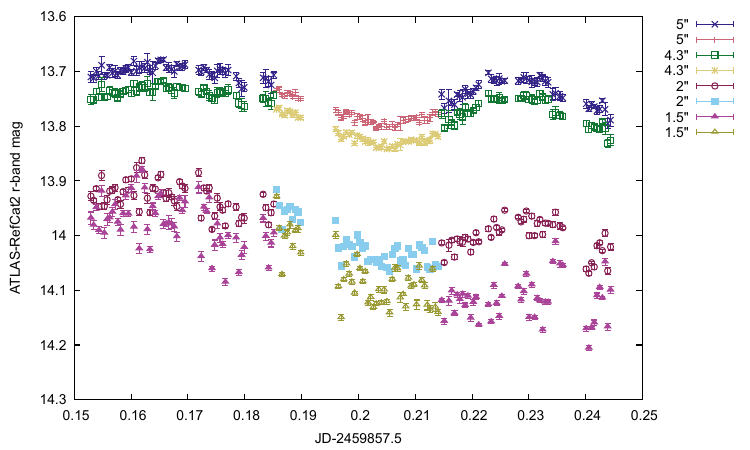}{0.5\textwidth}{(i) 5 October 2022}
\fig{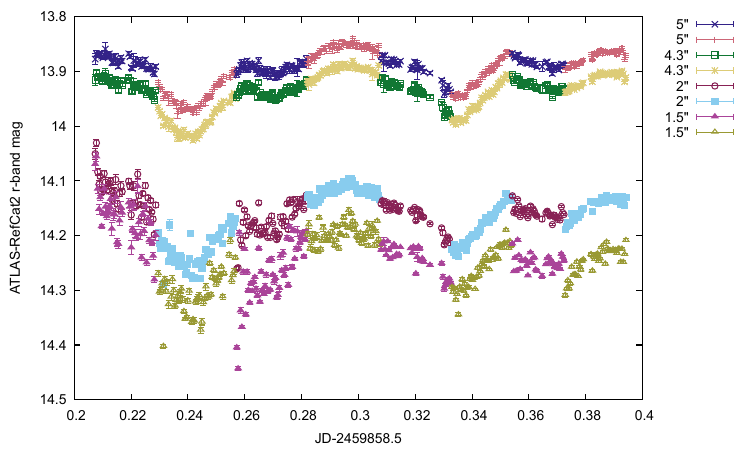}{0.5\textwidth}{(j) 6 October 2022}
}

\gridline{
\fig{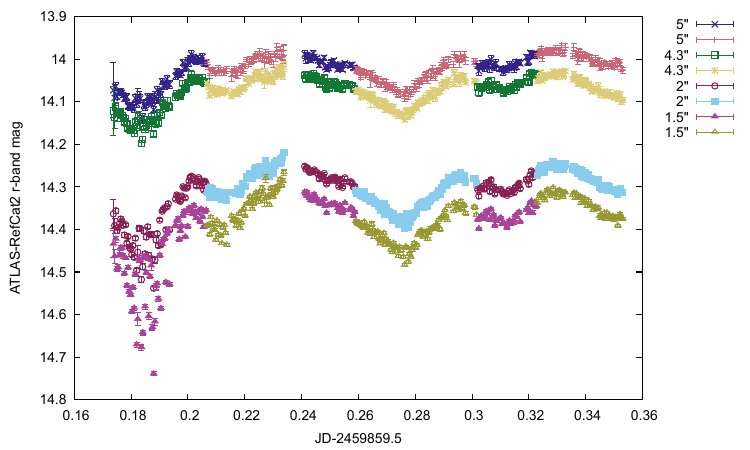}{0.5\textwidth}{(k) 7 October 2022}
\fig{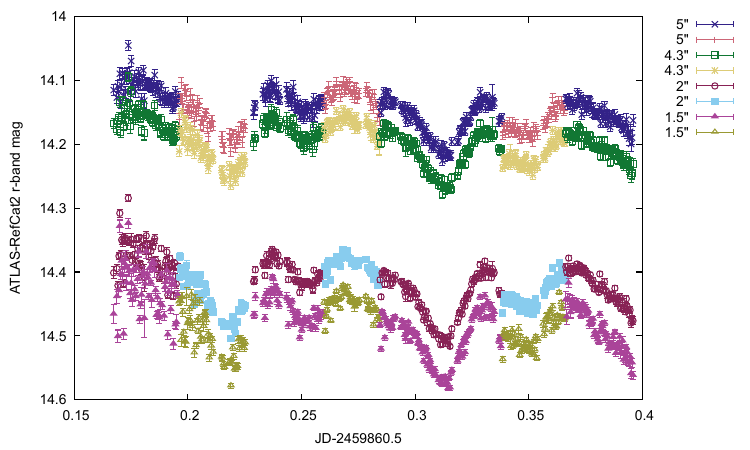}{0.5\textwidth}{(l) 8 October 2022}
}

\caption{[Continued]}
\end{figure}
\addtocounter{figure}{-1}
\begin{figure}

\gridline{\fig{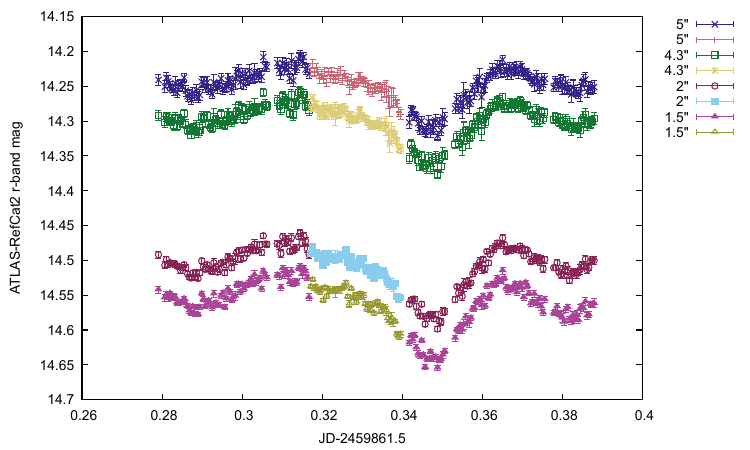}{0.5\textwidth}{(m) 9 October 2022}
\fig{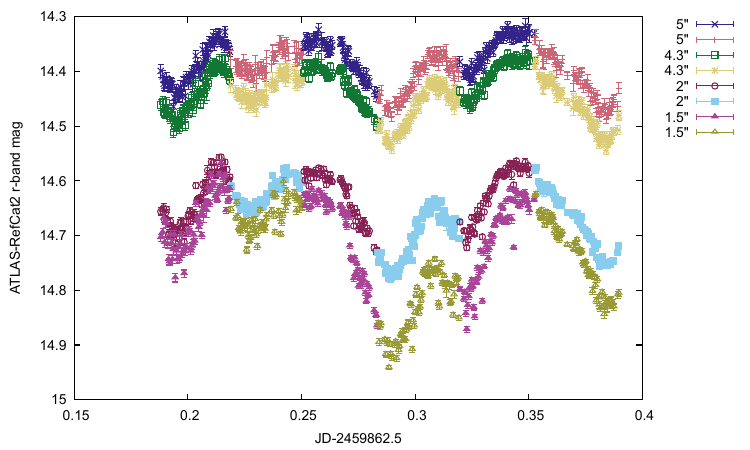}{0.5\textwidth}{(n) 10 October 2022}
}
\gridline{
\fig{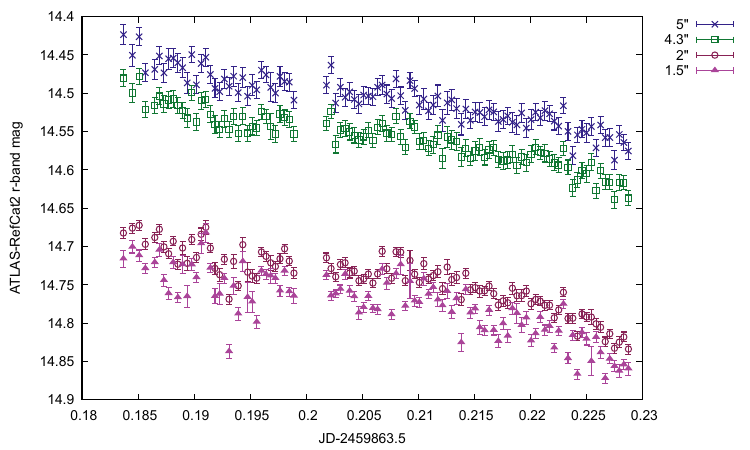}{0.5\textwidth}{(o) 11 October 2022}
\fig{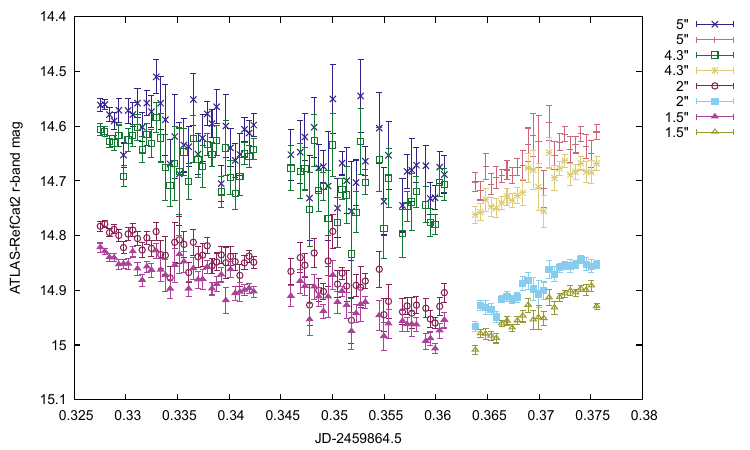}{0.5\textwidth}{(p) 12 October 2022}
}

\gridline{
\fig{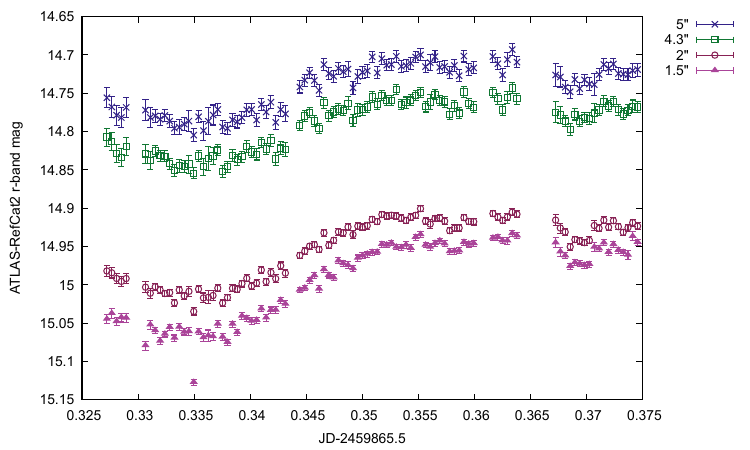}{0.5\textwidth}{(q) 13 October 2022}
\fig{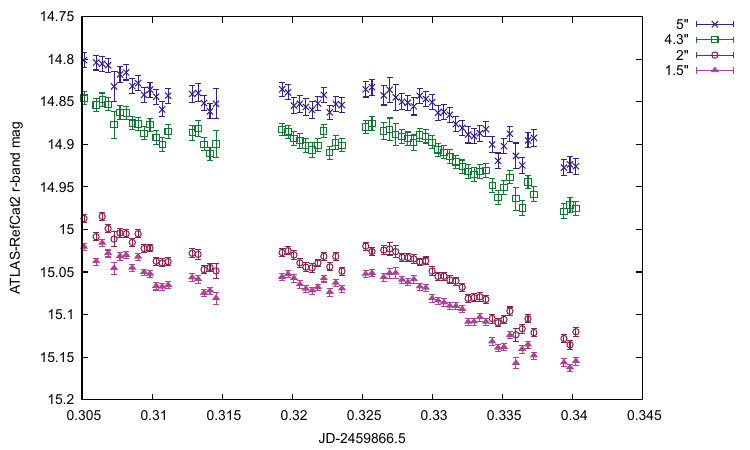}{0.5\textwidth}{(r) 14 October 2022}
}

\caption{[Continued]}
\end{figure}



\end{document}